\documentclass[a4paper,british,english,traditabstract]{aa}
\usepackage[T1]{fontenc}
\usepackage[latin9]{inputenc}
\usepackage{xcolor}
\usepackage{pdfcolmk}
\usepackage{verbatim}
\usepackage{longtable}
\usepackage{ifsym}
\usepackage{amsmath}
\usepackage{amssymb}
\usepackage{graphicx}
\PassOptionsToPackage{normalem}{ulem}
\usepackage{ulem}

\makeatletter

\pdfpageheight\paperheight
\pdfpagewidth\paperwidth

\newcommand{\noun}[1]{\textsc{#1}}
\providecommand{\tabularnewline}{\\}
\providecolor{lyxadded}{rgb}{0,0,1}
\providecolor{lyxdeleted}{rgb}{1,0,0}

\@ifundefined{definecolor}{\usepackage{color}}{}
\newcommand{\caii}{\ion{Ca}{ii}}
\newcommand{\citep}{\cite}
\newcommand{\citet}{\cite}
\def\kms{\rm km\,s^{-1}}
\authorrunning{Saviane et al.}

\def\W{\left\langle W^{\prime}\right\rangle}
\def\vvhb{$V-V_{\rm HB}$}

\usepackage{babel}
\usepackage{longtable}

\usepackage{babel}

\makeatother

\usepackage{babel}
\begin{document}

\title{Homogeneous Metallicities and Radial Velocities for Galactic Globular
Clusters%
\thanks{077.A-0775(A,B)%
}}

\subtitle{First CaT metallicities for twenty clusters}

\author{Ivo Saviane\inst{1}\and Gary S. Da Costa\inst{2} \and Enrico
V. Held\inst{3} \and Veronica Sommariva\inst{3,7} \and Marco Gullieuszik\inst{4}
\and Beatriz Barbuy\inst{5} \and Sergio Ortolani\inst{6} }

\institute{European Southern Observatory, Alonso de Cordova 3107, Santiago,
Chile\\
 \email{isaviane@eso.org}\and Research School of Astronomy \& Astrophysics,
Australian National University, Mt Stromlo Observatory, via Cotter
Rd, Weston, ACT 2611, Australia\\
 \email{gdc@mso.anu.edu.au} \and INAF, Osservatorio Astronomico
di Padova, vicolo Osservatorio 5, 35122 Padova, Italy\\
 \email{enrico.held@oapd.inaf.it} \and Royal Observatory, Avenue
Circulaire 3, 1180 Bruxelles, Belgium\\
 \email{M.Gullieuszik@oma.be}\and University of Sao Paulo, Rua do
Matao 1226, Sao Paulo 05508-900, Brazil\\
 \email{barbuy@astro.iag.usp.br} \and University of Padova, vicolo
Osservatorio 5, 35122 Padova, Italy\\
 \email{sergio.ortolani@unipd.it}\and INAF, Osservatorio Astrofisico
di Arcetri, Largo Enrico Fermi 5, 50125 Firenze, Italy \\
 \email{veronica@arcetri.astro.it}}

\offprints{isaviane@eso.org}

\date{Received ..., Accepted...}

\abstract{Well determined radial velocities and abundances are essential for
analyzing the properties of the Globular Cluster system of the Milky
Way. However more than 50\% of these clusters have no spectroscopic
measure of their metallicity. In this context, this work provides
new radial velocities and abundances for twenty Milky Way globular
clusters which lack or have poorly known values for these quantities.
The radial velocities and abundances are derived from spectra obtained
at the \caii\ triplet using the FORS2 imager and spectrograph at
the VLT, calibrated with spectra of red giants in a number of clusters
with well determined abundances. For about half of the clusters in
our sample we present significant revisions of the existing velocities
or abundances, or both. We also confirm the existence of a sizable
abundance spread in the globular cluster M54, which lies at the center
of the Sagittarius dwarf galaxy. In addition evidence is provided
for the existence of a small intrinsic internal abundance spread ($\sigma${[}Fe/H{]}$_{int}$
$\approx0.11$-$0.14$ dex, similar to that of M54) in the luminous
distant globular cluster NGC~5824. This cluster thus joins the small
number of Galactic globular clusters known to possess internal metallicity
({[}Fe/H{]}) spreads.}


\maketitle

\section{Introduction}

Galactic globular clusters (GGC) represent one of the fundamental
systems that allow a reconstruction of the early evolution of the
Milky Way: if their ages, metallicities and kinematics were known
with sufficient precision, then correlations of their kinematics and
chemical abundances with time would shed light on the dynamical and
chemical evolution of the protogalactic halo and bulge. To reach this
goal, large and homogeneous data samples are needed, and indeed considerable
progress has been seen in recent years. The largest collection of
color-magnitude diagrams (CMD) obtained with a single telescope (HST)
and uniform data reductions is that of Marín-Franch et al. (\cite{marin-franch_etal09},
MF09), where relative ages were calculated for $64$ clusters. With
respect to metallicities, Carretta et al. (\cite{carretta_etal09})
assembled a table of {[}Fe/H{]} values for $133$ clusters: the objects
were put on a single scale, but metallicities were computed based
on indices published in four different studies. Twenty~five clusters
were taken from the high resolution spectroscopic works of Carretta
\& Gratton (1997, CG97) and Kraft and Ivans (\cite{KI03}) -- which
have $15$ objects in common, and the rest come from Zinn \& West
(\cite{ZW84}, ZW84) and Rutledge et al. (1997b, R97). The $Q_{39}$
index of ZW84 is based on \emph{integrated light}, narrow-band imaging,
so the largest homogeneous spectroscopic study of individual globular
cluster stars is still that of R97. However their sample represents
only 44\% of the objects in the Harris (2010) \foreignlanguage{british}{catalogue},
which is a shortcoming for many investigations of the kind illustrated
above. For example $17$ clusters of the MF09 study ($\sim25\%$ of
the total) do not have an entry in R97, and their metallicity was
taken essentially from ZW84 (after a transformation to the CG97 scale).
Another example is offered by Forbes \& Bridges (\cite{forbes_bridges10})
who compiled ages and metallicities for $93$ GGCs, and based on their
age-metallicity relation (AMR) they estimated that $\sim25\%$ of
the clusters were accreted from a few dwarf galaxies. Most of their
table entries come from MF09, but for $29$ clusters ages and metallicities
had to be extracted from inhomogeneous sources. R97 based their work
on spectra obtained with the 2.5m Dupont telescope at Las Campanas,
so the absence of homogeneous data affects mostly outer-halo (or heavily
extincted) distant clusters, which is where relics of accretion are
most likely to be found.

To remedy this situation, we commenced a project to collect medium-resolution
spectra for clusters not included in the R97 sample. Spectra in the
$z$-band region of giant stars were obtained with FORS2 at the ESO/VLT
observatory to measure metallicities based on absorption lines of
the \caii\ triplet (CaT). We employ the ``reduced equivalent width''
method first introduced by Olszewski et al. (\cite{olszewski+91})
and Armandroff \& Da Costa (\cite{AD91}, hereafter AD91), which is
the same as used in R97. One caveat with this method is that $\alpha$-elements
(which include Ca) are enhanced in GGCs compared to the solar value,
with {[}$\alpha$/Fe{]}$\sim+0.3$ (e.g. Carney \cite{carney96}),
so it is important to identify clusters with anomalous enhancements,
especially when choosing the calibrators of the index-{[}Fe/H{]} relations.
For this reason, we also collected $V$-band spectra to explore a
metallicity ranking based on the strength of Fe and Mg lines. For
this spectral region, we are in the process of deriving effective
temperature $T_{{\rm eff}}$, gravity $\log g$ and metallicity {[}Fe/H{]}
from a full spectrum fitting of sample stars using a library of 1900
stars observed at high resolution with the ELODIE spectrograph, as
described in Katz et al. (\cite{katz+11}). The results will be presented
elsewhere (Dias et al., in preparation).

In a first paper based on our data (Da Costa et al. \cite{dacosta_etal09})
we discovered a metallicity spread in the giant stars of M22 almost
in parallel with the high resolution study of Marino et al. (\cite{marino+09}).
Here we publish the full catalog of CaT-based metallicities and radial
velocities, comprising $20$ clusters. As already discussed, the main
point of our project is the homogeneity of the data and analysis,
through which we add $\sim30\%$ more objects to the R97 sample%
\footnote{We had planned observations for $49$ clusters, but poor weather conditions
did not allow completion of the program.%
}. Furthermore, our calibration is based on a very large set of GC
templates accurately chosen to have well established abundances from
high-resolution studies.

A few more points can also be gathered from Table~\ref{tab:Observations-log},
where for each observed cluster we list the ``best'' source of metallicity
that we could find in the literature. For two clusters (NGC~6139
and NGC~6569) the metallicity is still based on the integrated-light
$Q_{39}$ index, and in the case of NGC~5824 it is based on the integrated-light
version of the CaT method (Armandroff \& Zinn \cite{AZ88}). Because
individual stars are not measured, for these clusters possible internal
metallicity spreads cannot be discovered, as the case of NGC~5824
clearly shows (see below). In nine cases high-resolution studies exist,
but almost all by different authors, causing concerns on the relative
abundance ranking. In addition high-resolution studies require large
observational and data analysis investments, limiting the number of
stars that can be measured and thus the statistical significance of
metallicity dispersions, if they are found. Finally six clusters have
{[}Fe/H{]} estimated via the equivalent width of Fe lines measured
on medium-resolution spectra, and calibrated on high-resolution spectra
of standard clusters. Again different methods and calibrations are
applied by different authors, raising concerns on the self-consistency
of the metallicity scales. To summarize, there are only two clusters,
where the same CaT method employed here was used (NGC~2808 and Pyxis)
to determine abundances. Hence we expect that the (reduced) EWs published
here will be the basis for all future abundance ranking of the clusters
listed in Table~\ref{tab:Observations-log}.

\section{Observations and reductions}

\begin{table}
\caption{Observations log \label{tab:Observations-log}}

\begin{centering}
\begin{tabular}{llllll}
 &  &  &  &  & \tabularnewline
\hline 
\hline 
Cluster  & UT  & $t_{{\rm exp}}$ (sec) & notes  & N  & {[}Fe/H{]}\tabularnewline
\hline 
 &  &  &  &  & \tabularnewline
\multicolumn{6}{c}{May 28, 2006}\tabularnewline
 &  &  &  &  & \tabularnewline
NGC3201  & 23:06  & 2$\times$7.5  & c  & 17  & \tabularnewline
M4  & 03:02  & 2$\times$1.9  & c  & 16  & \tabularnewline
M10  & 03:16  & 2$\times$20.9  & c  & 19  & \tabularnewline
NGC6397  & 05:31  & 2$\times$2.6  & c  & 17  & \tabularnewline
NGC6528  & 08:44  & 2$\times$61.2  & c  & 19  & \tabularnewline
NGC6553  & 09:04  & 2$\times$31.1  & c  & 18  & \tabularnewline
M71  & 09:23  & 2$\times$6.1  & c  & 14  & \tabularnewline
M15  & 09:39  & 2$\times$18  & c  & 18  & \tabularnewline
 &  &  &  &  & \tabularnewline
Rup106  & 23:22  & 2$\times$349  & 2nd & 15  & {\scriptsize h/BWZ97}\tabularnewline
NGC2808  & 00:14  & 2$\times$16.8  &  & 19  & {\scriptsize CaT/R97}\tabularnewline
NGC5824  & 00:35  & 2$\times$248.9  &  & 19  & {\scriptsize i/AZ88}\tabularnewline
Lynga7  & 01:26  & 2$\times$200.2  &  & 19  & {\scriptsize Fe/TF95}\tabularnewline
NGC6139  & 03:34  & 2$\times$264.4  &  & 19  & {\scriptsize i/Q39}\tabularnewline
Terzan3  & 03:55  & 2$\times$230.5  &  & 19  & {\scriptsize Fe/C99}\tabularnewline
NGC6325  & 04:18  & 2$\times$214.1  & b  & 19  & {\scriptsize Fe/M95a}\tabularnewline
NGC6356  & 04:36  & 2$\times$69.4  & b  & 19  & {\scriptsize Fe/M95a}\tabularnewline
NGC6380  & 04:59  & 2$\times$615.6  &  & 19  & {\scriptsize Fe/C99}\tabularnewline
NGC6440  & 05:48  & 2$\times$295.4  & b  & 19  & {\scriptsize h/OVR08}\tabularnewline
NGC6441  & 06:37  & 2$\times$96  & b  & 19  & {\scriptsize h/OVR08}\tabularnewline
NGC6558  & 07:03  & 2$\times$60.7  & b  & 19  & {\scriptsize h/B+07}\tabularnewline
Pal7  & 07:25  & 2$\times$97.1  &  & 18  & {\scriptsize Fe/C99}\tabularnewline
NGC6569  & 07:51  & 2$\times$88.3  &  & 19  & {\scriptsize i/Q39}\tabularnewline
M22  & 08:14  & 2$\times$13.3  & m. 1  & 19  & {\scriptsize h/BW92/-}\tabularnewline
M22  & 08:28  & 2$\times$13.3  & m. 2  & 19  & {\scriptsize -/LBC91/M+09}\tabularnewline
M22  & 09:55  & 2$\times$13.3  & m. 3  & 18  & \tabularnewline
 &  &  &  &  & \tabularnewline
\multicolumn{6}{c}{May 29, 2006}\tabularnewline
 &  &  &  &  & \tabularnewline
Pyxis  & 23:23  & 2$\times$1258.9  & 2nd & 18  & {\scriptsize CaT/PKM00}\tabularnewline
HP1  & 05:31  & 2$\times$488.1  & b, X  & 39  & {\scriptsize h/B+06}\tabularnewline
M54  & 08:15  & 2$\times$278.8  &  & 19  & {\scriptsize h/BWG99/-}\tabularnewline
 &  &  &  &  & {\scriptsize -/B+08/Car10}\tabularnewline
Terzan7  & 08:53  & 2$\times$334.9  & sag, X  & 25  & {\scriptsize h/T+04/S+05}\tabularnewline
NGC7006  & 09:20  & 2$\times$680  & 2nd, X  & 25  & {\scriptsize h/K+98}\tabularnewline
 &  &  &  &  & \tabularnewline
\hline 
\end{tabular}
\par\end{centering}

\tablefoot{N is the number of stars observed in each cluster. c=calibration,
X=done with mask (MXU) to increase slitlets density, b=bulge, sag=Sagittarius
dSph, 2nd=second parameter. The metallicity source is coded as method/references,
where the reference acronym can be found in the bibliography list.
The method can be (h)igh resolution, (i)ntegrated light, medium-resolution
EW of (Fe) lines, or (CaT) method.} 
\end{table}

\begin{figure}
\begin{centering}
\includegraphics[width=0.9\columnwidth]{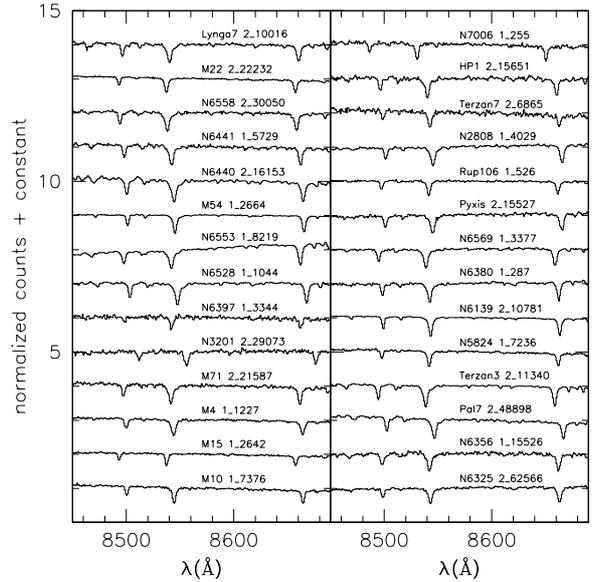} 
\par\end{centering}

\caption{A selection of typical fully reduced spectra is shown, one for each
of the $28$ clusters observed. For each spectrum the cluster and
star identification are given above the continuum. \label{fig:A-random-selection}}
\end{figure}

\subsection{Selection of targets}

The core target list was assembled as follows. We started from the
catalog of GGCs published in Harris (\cite{harris96}, version Feb.
2003, hereafter H96) and considered all objects visible during ESO
P77, i.e. having RA between $12$~h and $0$~h and DEC southern
of $+20^{\circ}$ (which makes them observable from Paranal for at
least 2hr at airmass $X<1.5$). We also required to reach at least
$M_{V}=-1$ along the RGB (see AD91), and to have $V=20$ as the faintest
RGB star to limit exposure times to $\sim1$~hr. Further, to exclude
the closest clusters, a requirement $(m-M)_{V}>16.75$ was also imposed.
Finally we used the compilations of Pritzl et al. (\cite{PVI05})
and Gratton et al. (\cite{gratton+04}) to remove those clusters that
had CaT measurements and to add calibration clusters. A few objects
that we deemed interesting for a number of reasons were then added
back to the list:Terzan~7 of the Sagittarius dwarf spheroidal galaxy,
NGC7006 as `second parameter' object, and NGC6325, NGC6356, NGC6440,
NGC6441, and HP1 as bulge members. The final list eventually had $57$
entries, including the $8$ calibrators.

\subsection{Observations}

Our data were obtained with FORS2 (Appenzeller et al. \cite{appenzeller+98}),
working at the Cassegrain focus of VLT/UT1-Antu. The instrument offers
two ways of realizing multi-object spectroscopy, either with $19$
movable slitlets (MOS mode), or by fabricating masks to be inserted
in a mask exchange unit (MXU mode). To define the masks and MOS slitlets
configurations, nine hours of pre-imaging in the $V$ and $I$ bands
were used to observe the $49$ main targets in service mode, while
pre-imaging for the calibrators was available from previous VLT runs.
The $6\farcm8\times6\farcm8$ field-of-view was centered on the cluster
for the sparser or more distant systems, but was offset from the centers
for the nearer and richer clusters. For each cluster the resulting
color-magnitude diagram was used to select targets on the RGB. Afterward,
spectra were collected in visitor mode, during a two-night run at
the end of May 2006. The choice of MXU or MOS mode was based on the
stellar density. The CaT spectral region was covered with the 1028z+29
grism and the OG590+32 order-blocking filter, yielding a maximum spectral
coverage of $\sim$7700\AA{}--9500\AA{}\/ at a scale of 0.85\AA{}\/
per (binned) pixel. In all cases the slit width was defined at $1\arcsec$.
For the MOS mode the number of slits is sometimes smaller than $19$,
because of the targets' distribution on the sky (see Table~\ref{tab:Observations-log}).
As the table shows, in three cases masks were used to cover higher
density clusters, with up to 39 slitlets per mask. The target star
magnitudes typically cover $\sim$3 mag along the RGB, and two exposures
were obtained to allow removal of cosmic-rays. Because of the limited
time and bad weather conditions, spectra could be collected for $20$
clusters only (plus the eight calibrators). The observing log is reported
in Table~\ref{tab:Observations-log}: 23 clusters could be observed
in the first night, while clouds in the second night limited observations
to 5 clusters only.

\subsection{Extraction of spectra}

All spectra were extracted using the FORS2 pipeline version 1.2 (Izzo
\& Larsen \cite{izzo_larsen08}), which splits the processing into
two steps. First, using daytime calibration frames, slit positions
are found, and the wavelength calibration and distortion map for each
slit are created. Then the products of the first step are used to
reduce the target spectra. The pipeline logs give a mean residual
scatter around the wavelength calibration of $0.24$ pixels (i.e.
$0.2$~\AA{}\ or $\sim7\,\kms$ for a dispersion of 0.82~\AA{}~px$^{-1}$),
a mean spectral resolution of $\approx2440$, and a mean \textsc{fwhm}
of the arc lines of $3.51\pm0.07$~\AA{} (130~$\kms$). Given this
resolution we expect individual radial velocities to have uncertainties
of 10-15~$\kms$, and thus the mean velocity for a cluster will have
an uncertainty 3 to 4 times smaller depending on the number of members.
This is consistent with the uncertainties given in Table~\ref{tab:Radial-Velocities}.
We note though that a comparison of our mean cluster velocities with
well determined values in the literature (see Sect.~\ref{sub:Comparison-with-Harris})
suggests that the true uncertainties in our mean velocities are actually
somewhat larger, probably as a result of less easily quantified systematic
effects such as mask-centering errors, etc.

The software corrects for bias and flatfield, and computes a local
sky background for each slit, to be subtracted from the object spectra.
The Horne (\cite{horne86}) optimal extraction is applied. The spectra
are normalized by exposure time, and the wavelength solution is aligned
to a reference set of $>$20 sky lines by applying an offset (0.09~px
on average). The two exposures were average-combined after pipeline
processing. The S/N ratios for the final spectra varied from $\sim$110
for the brightest stars to $\sim$25 for the faintest in a typical
exposure. To illustrate the quality of our data, Fig.~\ref{fig:A-random-selection}
presents one spectrum for each of the observed clusters.

\section{Radial Velocities}

\begin{table}
\caption{Radial Velocities \label{tab:Radial-Velocities}}

\noindent \begin{centering}
\begin{tabular}{lrrlrrr}
 &  &  &  &  &  & \tabularnewline
\hline 
\hline 
Cluster  & N  & \multicolumn{2}{c}{$\left\langle v_{{\rm R}}\right\rangle \pm$err} & \multicolumn{2}{c}{$\left\langle v_{{\rm R}}\right\rangle \pm$err} & $\Delta v_{{\rm R}}$\tablefootmark{a}\tabularnewline
 &  & \multicolumn{2}{c}{$\kms$} & \multicolumn{2}{c}{$\kms$} & $\kms$\tabularnewline
\hline 
 &  &  &  &  &  & \tabularnewline
 &  &  &  & \multicolumn{2}{c}{H10} & \tabularnewline
\hline 
 &  &  &  &  &  & \tabularnewline
\multicolumn{7}{l}{a) metal-poor standard clusters}\tabularnewline
 &  &  &  &  &  & \tabularnewline
M4  & 15  & 72  & 6  & 70.7  & 0.2  & 1.3\tabularnewline
M10  & 13  & 86  & 2  & 75.2  & 0.7  & 10.8\tabularnewline
NGC6397  & 16  & 12  & 3  & 18.8  & 0.1  & $-$6.8\tabularnewline
M71  & 11  & $-$25  & 4  & $-$22.8  & 0.2  & $-$2.2\tabularnewline
M15  & 18  & $-$111  & 5  & $-$107.2  & 0.2  & $-$4.0\tabularnewline
NGC3201  & 17  & 486  & 4  & 494.0  & 0.2  & $-$8.0\tabularnewline
 &  &  &  &  &  & \tabularnewline
 &  &  &  &  &  & \tabularnewline
\multicolumn{7}{l}{b) metal$-$poor program clusters }\tabularnewline
 &  &  &  &  &  & \tabularnewline
Pyxis  & 8  & 42  & 4  & 34.3  & 1.9  & 7.6\tabularnewline
NGC2808  & 17  & 91  & 4  & 101.6  & 0.7  & $-$10.6\tabularnewline
Rup106  & 9  & $-$39  & 4  & $-$44.0  & 3.0  & 5.0\tabularnewline
NGC5824  & 17  & $-$11  & 2  & $-$27.5  & 1.5  & 16.5\tabularnewline
Lynga 7  & 9  & 22  & 3  & 8.0  & 5.0  & 8.0\tabularnewline
NGC6139  & 15  & 34  & 4  & 6.7  & 6.0  & 27.3\tabularnewline
Terzan 3  & 13  & $-$131  & 4  & $-$136.3  & 0.7  & 5.3\tabularnewline
NGC6325  & 10  & 28  & 2  & 29.8  & 1.8  & $-$1.8\tabularnewline
HP1\tablefootmark{b}  & 8  & 41  & 4  & 45.8  & 0.7  & $-$4.8\tabularnewline
NGC6558  & 5  & $-$198  & 4  & $-$197.2  & 1.6  & $-$0.8\tabularnewline
Pal 7  & 14  & 164  & 3  & 155.7  & 1.3  & 8.3\tabularnewline
NGC6569  & 7  & $-$47  & 3  & $-$28.1  & 5.6  & $-$18.9\tabularnewline
M22  & 51  & $-$150  & 2  & $-$146.3  & 0.2  & $-$3.7\tabularnewline
M54  & 19  & 137  & 2  & 141.9  & 0.5  & $-$4.3\tabularnewline
NGC7006  & 20  & $-$379  & 1  & $-$384.1  & 0.4  & 5.1\tabularnewline
 &  &  &  &  &  & \tabularnewline
\multicolumn{7}{l}{c) metal$-$rich standard clusters }\tabularnewline
 &  &  &  &  &  & \tabularnewline
NGC6528  & 4  & 205  & 2  & 206.6  & 1.4  & $-$1.6\tabularnewline
NGC6553  & 18  & $-$9  & 4  & $-$3.2  & 1.5  & $-$5.8\tabularnewline
 &  &  &  &  &  & \tabularnewline
\multicolumn{7}{l}{d) metal$-$rich program clusters }\tabularnewline
 &  &  &  &  &  & \tabularnewline
NGC6356  & 12  & 67  & 4  & 27.0  & 4.0  & 39.0\tabularnewline
NGC6380  & 9  & 7  & 3  & $-$3.6  & 2.5  & 10.6\tabularnewline
NGC6440  & 8  & $-$76  & 4  & $-$76.6  & 2.7  & 1.6\tabularnewline
NGC6441  & 7  & 18  & 4  & 16.5  & 1.0  & 1.5\tabularnewline
Ter7  & 14  & 154  & 3  & 166.0  & 4.0  & $-$12.0\tabularnewline
 &  &  &  &  &  & \tabularnewline
\hline 
\end{tabular}
\par\end{centering}

\tablefoot{\tablefoottext{a}{$\Delta v_{{\rm R}}$ is in the
sense FORS2-H10.} \tablefoottext{b}{HP1: The difference listed
is with respect to the heliocentric velocity given by Barbuy et al
(2006), which we give in lieu of the Harris catalog value. It appears
the catalog lists the uncorrected observed value.}} 
\end{table}

\begin{figure}
\begin{centering}
\includegraphics[angle=-90,width=0.9\columnwidth]{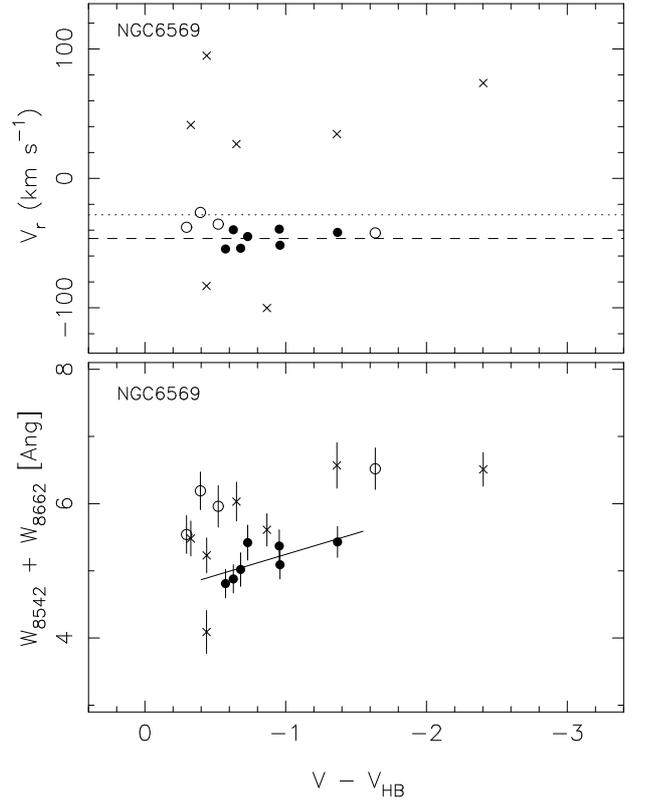} 
\par\end{centering}

\caption{Membership selection for stars observed in NGC6569. The upper panel
shows the heliocentric radial velocity plotted against $V-V_{{\rm HB}}$.
The dotted line is the cluster velocity given by the Harris \foreignlanguage{british}{catalogue},
the dashed line is the mean velocity for the 7 stars ultimately selected
as cluster members. The stars unambiguously identified as radial velocity
non-members are plotted as x-signs. The lower panel shows the sum
of the strengths of the $\lambda$8542 and $\lambda$8662 lines plotted
against $V-V_{{\rm HB}}$. Here, 4 stars, plotted as open circles,
have line strengths incompatible with the remaining 7, plotted as
filled circles, which have a small dispersion about a line of slope
-0.627 \AA{}~mag$^{-1}$ (see Sect.~\ref{sec:Metallicities}). These
4 stars are classified as line-strength non-members. \label{fig:Membership-selection-for}}
\end{figure}

Heliocentric radial velocities of the single stars were computed with
the \texttt{rvidlines} task in \noun{iraf}'s \texttt{rv} package using
all three of the \caii\ triplet lines. Afterward, the cluster velocity
was computed by averaging the velocities of the cluster members. In
doing so, stars with significantly discrepant line strengths were
excluded even if they had velocities compatible with cluster membership.
The velocity errors were computed as standard errors of the mean as
determined from the standard deviation of the individual velocities.
Our results are summarized in Table~\ref{tab:Radial-Velocities}
where N is the number of cluster members used in calculating the mean
velocity.

In detail our assessment of cluster membership for each star observed
was based on two assumptions: that the range in the radial velocities
of member stars was small, and that the dispersion in the measured
equivalent widths about the fitted line was comparable to the measurement
errors. The latter assumption is equivalent to assuming the intrinsic
abundance dispersion in a cluster is small, an assumption that despite
recent discoveries (e.g., Sec.~\ref{sub:An-abundance-spread/5824})
remains generally valid: most globular clusters are mono-metallic.
Nevertheless, careful consideration was given to the likely membership
in every case where an observed star had a velocity consistent with
that of the cluster but a line strength different from that expected
for its $V-V_{{\rm HB}}$ value, given the line strengths of other
candidate members. For the more luminous stars in the more metal-rich
clusters a discrepant weaker line strength was often the result of
a depressed pseudo-continuum caused by the $\sim$8440~\AA{} TiO
bandheads (see Olzsewski et al. \cite{olszewski+91}). Such occurrences
were readily identifiable and the stars showing TiO were included
in the cluster velocity calculation, if the velocity was consistent
with other members, but not in the determination of the cluster line
strengths. The final selection of members was also required to define
a sensible sequence in the color-magnitude diagram derived from the
cluster pre-imaging, though given the process by which the stars to
observe were initially selected, this provides only a consistency
check. We also note that the spectra of every star observed was visually
inspected so that any effects from the occasional poorly removed cosmic
ray or inadequate sky-subtraction could be allowed for.

We illustrate our membership selection process in Fig.~\ref{fig:Membership-selection-for}
for the cluster NGC6569 which is typical of the sample of program
clusters. The upper panel shows the measured radial velocities for
18 of the 19 stars observed in this cluster. The spectrum of the remaining
star had an instrumental defect in the vicinity of the $\lambda$8662
line of the \caii\ triplet and so was not used in the subsequent
analysis. The figure shows that there are 7 stars whose velocities
differ significantly from the others; such stars are unambiguously
non-members. The lower panel shows the sum of the strengths of the
$\lambda$8542 and $\lambda$8662 lines of the \caii\ triplet plotted
against $V-V_{{\rm HB}}$. Here four of the stars with velocities
compatible with cluster membership have line strengths that are significantly
larger than those for the remaining seven, whose line strengths are
consistent with measurement error driven scatter about a single (line
strength, $V-V_{{\rm HB}}$) relation. Consequently, we consider these
4 stars as line strength non-members - the alternative that there
is a second population in the cluster with metallicity $\sim$0.8~dex
higher is much less plausible.

We also verified that our velocity results are not affected by the
choice to measure the line centers of the three \caii\ triplet lines
rather than using cross-correlation techniques. Specifically, we reanalyzed
the velocities for the standard cluster NGC~3201 and the program
clusters NGC~5824 and HP1 by applying the \texttt{fxcor} task within
\noun{iraf}. The spectral region correlated was 8400-8770~\AA{} which
is largely free of atmospheric absorption contamination and which
contains a number of weak stellar lines. For NGC~3201 and NGC~5824
the template employed was a cluster member while for HP1 the template
used was M10 star 1\_3150. The mean difference between the velocities
from the cross-correlation process and from the triplet line center
measurements was $0.8\,\kms$, $0.8\,\kms$ and $1.0\,\kms$ for NGC~3201,
NGC~5824 and HP1, respectively. The standard deviations and the number
of stars used were $5.6\,\kms$ (16), $1.8\,\kms$ (17) and $4.1\,\kms$
(26), respectively. For 7 HP1 member stars the mean difference and
standard deviation were $1.6\,\kms$ and $3.9\,\kms$, respectively.
We conclude that our adopted velocities and their uncertainties are
not affected by the choice of measurement technique.

\subsection{Comparison with Harris catalog values \label{sub:Comparison-with-Harris}}

\begin{figure}
\begin{centering}
\includegraphics[angle=-90,width=0.9\columnwidth]{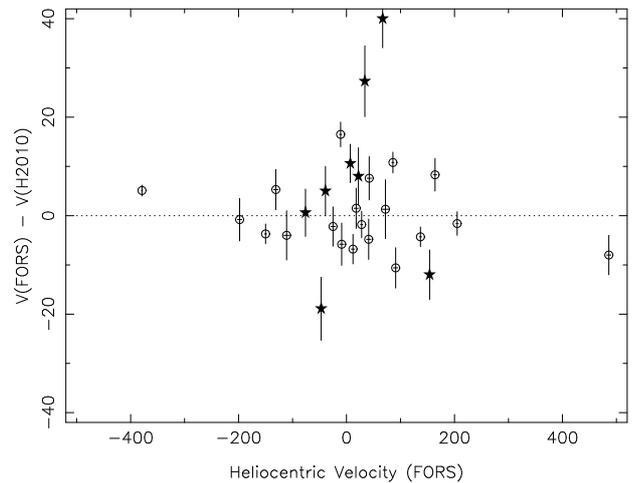} 
\par\end{centering}

\caption{The velocity differences between our values and those of H10 are plotted
here versus our velocities. The open circles are for clusters where
Harris lists an error of less than $2\,{\rm km\, s}^{-1}$ and filled
stars are for clusters where the H10 listed error exceeds $2\,{\rm km\, s}^{-1}$.
The vertical error bars are given by the quadratic sum of our and
H10 errors. \label{fig:The-velocity-differences}}
\end{figure}

To verify the \foreignlanguage{british}{zero-point} of our velocities,
we first compared our velocities with those from H96 (further updated
in December 2010, hereafter H10) using only the clusters with a listed
uncertainty of less than $2\,\kms$. There are $20$ such clusters,
and for these the mean difference is $V_{{\rm FORS}}-V_{{\rm H10}}=0.10\,\kms$
with a standard deviation of $7~\kms$. This indicates the true uncertainties
in our mean velocities are likely of order $5$-$6\,\kms$ rather
than $3$-$4\,\kms$, the mean of the errors listed in Table~\ref{tab:Radial-Velocities}.
Nevertheless, the effectively zero mean offset indicates excellent
consistency between our determinations and existing well determined
ones. The velocity differences $V_{{\rm FORS}}-V_{{\rm H10}}$ for
these clusters are shown as open circles in Fig.~\ref{fig:The-velocity-differences}.

The largest difference among these clusters is for NGC~5824 where
we find a velocity of $-11\pm2~\kms$ from $17$ members while the
Harris catalog lists a velocity of $-27.5\pm1.5~\kms$. The catalog
entry is essentially that of Dubath, Meylan \& Mayor (\cite{dubath+97})
who give $V_{{\rm r}}=-26.0\pm1.6\,\kms$ from an integrated spectrum
of the cluster \foreignlanguage{british}{centre}. Earlier less precise
values, e.g. $-$30~$\kms$ (Armandroff \& Zinn \cite{AZ88}), $-$28~$\kms$
(Zinn \& West \cite{ZW84}) and $-$38~$\kms$ (Hesser, Shawl \&
Meyer \cite{hesser+86}; hereafter HSM86) are not inconsistent with
the Dubath et al. (\cite{dubath+97}) value. This suggests that our
NGC5824 velocities might be all systematically $\sim$15-20~$\kms$
too high. Such offset represents 1/8 to 1/6 of the FWHM resolution
element, and could be due to an overall mis-centering of the MOS mask.
This interpretation is supported by a comparison of our velocities
with those determined from high resolution spectroscopy of three NGC5824
stars by Villanova and Geisler (private communication). The mean velocity
found from the high dispersion spectra is $-$30.8$\pm$2.4~$\kms$
while for the same three stars our mean velocity is $-$9$\pm$3~$\kms$,
supporting the idea of a systematic offset in the FORS2 velocities.
We emphasize though that the existence of this possible velocity offset
does not affect in any way the classification of the 17~FORS2 stars
as NGC 5824 members. We intend to obtain additional spectra of stars
in this cluster, which will either confirm or lead to a revision of
our value.

For the remaining eight clusters in our sample, the uncertainty listed
in the H10 \foreignlanguage{british}{catalogue} exceeds $2~\kms$.
The difference between our new determinations and the \foreignlanguage{british}{catalogue}
values are plotted as filled stars in Fig.~\ref{fig:The-velocity-differences}.
For five of the clusters the previous estimates are not inconsistent
with our newer and better established values, but for the remaining
three (NGC~6139, NGC~6569 and NGC~6356) the discrepancy with the
previous estimates is substantial. For NGC~6139 and NGC~6569 the
only previous determinations come from relatively low resolution spectra.
In particular, for NGC~6139 the previous determinations are $8\pm7~\kms$
from Webbink (\cite{webbink81}) and $4\pm12~\kms$ from HSM86. Our
value of $34\pm4~\kms$ is undoubtedly preferable. For NGC~6569 our
determination ($-47\pm3~\kms$) is in reasonable accord with that
given by Zinn \& West (\cite{ZW84}), $-36\pm14~\kms$, but disagrees
with that, $-26\pm6~\kms$ given by HSM86. Again our new determination
is to be preferred.

The largest discrepancy is for NGC~6356 where we find a velocity
of $66\pm4~\kms$ from $13$ member stars while the catalog of H10
lists a velocity of $27\pm4~\kms$. Our value is supported by the
value given by Minniti (\cite{minniti95}) of $56\pm5~\kms$. It is
possible that in generating the catalog entry the (labeled uncertain)
HSM86 velocity has been over weighted and the Minniti (1995) value
under weighted. Regardless of the origin of the discrepancy it is
likely that our value is preferable.

Finally Terzan 7 and HP1 deserve further comments. The velocity quoted
by H10 for the first cluster is that of DA95 ($4$ stars), which is
$12\,\kms$ higher than our value. However Sbordone et al. (\cite{sbordone+05})
have published velocities for $5$ stars that are in better agreement
with our result. Averaging the numbers from their Table~1, the velocity
is $159\pm1\,\kms$, with a difference of only $+5\,\kms$ compared
to us. In the case of HP1, it appears that the velocity quoted in
H10 is the geocentric value given in Barbuy et al. (\cite{barbuy+06}),
while the heliocentric value agrees with our result within the errors.
Indeed that velocity is the one given in the H10 column of Table~\ref{tab:Radial-Velocities}.

\section{Metallicities \label{sec:Metallicities}}

\begin{figure*}
\begin{centering}
\includegraphics[angle=-90,width=0.7\textwidth]{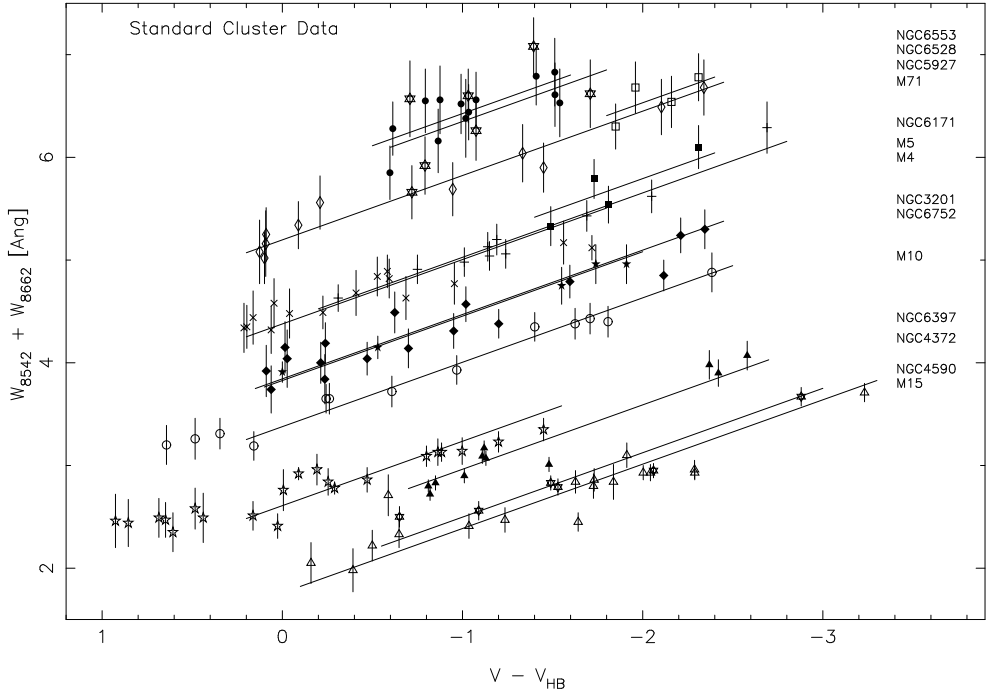} 
\par\end{centering}

\caption{Plot of \caii\ line strength ($W_{8542}+W_{8662}$) against magnitude
difference from the horizontal branch ($V-V_{HB}$) for the standard
clusters. In order of increasing $W_{8542}+W_{8662}$ values at $V-V_{HB}$
= --1.5, the solid lines are for clusters M15 (individual stars plotted
as open 6-pt star symbols), NGC~4590 (open triangles), NGC 4372 (filled
triangles), NGC 6397 (open 5-pt stars), M10 (open circles), NGC 6752
(filled 5-pt stars), NGC 3201 (filled diamonds), M4 (x-symbols), M5
(plus symbols), NGC 6171 (filled squares), M71 (open diamonds), NGC
5927 (open squares), NGC 6528 (6-pt star symbols), and NGC 6553 (filled
circles). The data for each cluster has been fit with a line of slope
$-0.627$\AA{}/mag for $V-V_{HB}$ $\leq$ 0.2. Vertical bars on each
point show the measurement uncertainty in the line strengths. \label{fig:cal_clusters_ew_fig-1}}
\end{figure*}

Metallicities were computed by following the well established method
originally proposed by AD91. Initially, for each star in all clusters
the equivalent widths of the two strongest CaT lines ($\lambda8542$,
$\lambda8662$) are measured, and their sum $\Sigma W$ is computed.
Then using only calibration clusters, $\Sigma W$ values are plotted
vs. $V-V_{{\rm HB}}$, and the slope of the $\Sigma W$ vs. $V-V_{{\rm HB}}$
relation is found (see Fig.~\ref{fig:cal_clusters_ew_fig-1}). To
a good approximation the slope $a$ is independent from metallicity
within the range spanned by GGCs, so it can be used to define so called
reduced equivalent widths $W^{\prime}=W_{8542}+W_{8662}-a\,(V-V_{{\rm HB}})$
for all stars in each cluster. In this way the gravity dependence
of EWs is empirically removed, and an average value $\left\langle W^{\prime}\right\rangle $
can be computed%
\footnote{$\W$ also coincides with the intercept of the linear fit to the $(\Sigma W,$$V-V_{{\rm HB}})$
points%
}. Finally the {[}Fe/H{]} vs. $\left\langle W^{\prime}\right\rangle $
relation for the calibration clusters was fit with a low-order polynomial
to define the calibration relation (see Fig.~\ref{fig:Same-as-Fig.,-1-1}).
At this point $W^{\prime}$ values can be computed for each star in
the rest of the clusters, and they can be converted into {[}Fe/H{]}
values by means of the relation found above.

To obtain a more reliable abundance calibration, the standard cluster
observations from Gullieuszik et al. (\cite{MG09}, G09) were also
analyzed. These spectra were obtained with the same instrument setup
and were reduced using the same methods employed here, the only difference
being the slit width ($0\farcs8$ for G09 and $1\arcsec$ here). The
additional calibration clusters from G09 are NGC~4590, NGC~4372,
NGC~5927, NGC~6397 (no stars in common with our data set), NGC~6528
(no stars in common with our dataset), NGC~6752, M5 and NGC~6171.
The total of calibration clusters is then 14 (see Table~\ref{tab:Input-data-for-1}).

It should be noted that the CaT method works on the assumption that
$V_{{\rm HB}}$ depends mainly on {[}Fe/H{]}, and that other stellar
parameters play a secondary role. While this is true for the relative
ages of old stellar systems like globular clusters%
\footnote{The effect of age on the CaT method is discussed in Da Costa \& Hatzidimitriou
(\cite{dacosta-hatzidimitriou-98}), Cole et al. (\cite{cole_etal04}),
Pont et al. (\cite{pont+2004}), and Carrera et al. (\cite{carrera+07}).%
}, cluster-to-cluster differences in helium abundance might instead
be significant, and might introduce additional scatter in the metallicities
yielded by the method. In general GGCs share a common He abundance
(Buzzoni et al. \cite{buzzoni+83}, Zoccali et al. \cite{zoccali+00},
Cassisi et al. \cite{cassisi+03}), but things might be different
for bulge clusters. Recently Nataf et al. (\cite{nataf+11}) explained
the difference between the luminosity of the RGB bump of the Galactic
bulge, and that predicted by the luminosity-metallicity relation of
Galactic globular clusters, by postulating that bulge stars have an
He enhancement $\Delta Y=0.06$ (see also Renzini \cite{renzini94}).
To see the effect of such enhancement on the metallicity computed
via the CaT method, one would need to take stellar models computed
with different helium abundances, and then compute EWs of Ca lines
and $V$-band luminosities using atmospheres with gravities and effective
temperatures that are changed accordingly. Indeed we plan to carry
out these tests in a forthcoming paper, while here we can check what
is the effect on the luminosity of the HB of the He enhancement quoted
above. According to Renzini (\cite{renzini77}) the bolometric luminosity
of the HB varies as $\Delta M_{{\rm HB}}=-4.7\times\Delta Y$ at a
fixed metallicity and age. The bulge HB, and presumably that of its
clusters, might therefore be brighter by $\Delta V_{{\rm HB}}\sim0.28$~mag.
Such shift in luminosity would shift $\W$ by $0.28\times0.627=0.18$~\AA{},
where $0.627$ is the slope of the $\Sigma W,V-V_{{\rm HB}}$ relation
(Fig.~\ref{fig:cal_clusters_ew_fig-1}). The change in $\W$ causes
a change in {[}Fe/H{]} that depends on the metallicity itself (Fig.~\ref{fig:Same-as-Fig.,-1-1}),
being $\sim0.05$~dex for metal-poor clusters, and up to $\sim0.2$~dex
for metal-rich ones. However bulge clusters in Fig.~\ref{fig:Differences-in-metallicity}
have very small {[}Fe/H{]} differences with respect to the literature,
and in line with those of the rest of the clusters. An exception is
HP1, but the difference with the literature is not systematic: our
{[}Fe/H{]} is smaller than that of H10 and larger than that of C09/Appendix1
by the same $\sim0.3$ dex, which is more than what an He enhancement
would predict. The conclusion is that the cluster-to-cluster scatter
in He content should not affect the results of this work.

In the following sections the analysis is described in more detail,
starting with a description of how photometric catalogs and equivalent
widths were obtained, and then moving to the {[}Fe/H{]}-$\left\langle W^{\prime}\right\rangle $
calibration relation.

\subsection{Photometry}

The pre-imaging data (which included both short and long exposures)
were used to create photometric catalogs. Stetson's \noun{daophot/allstar}
package (Stetson \cite{pbs87}, \cite{pbs94}) was used to carry out
the Point-Spread-Function photometry. Instrumental magnitudes were
calibrated by using color terms and zero points provided by ESO as
part of their routine quality control%
\footnote{see http://www.eso.org/observing/dfo/quality/FORS2/qc/\\
 /photcoeff/photcoeffs\_fors2.html%
}. We estimate that the zero-points on the standard $V$, $I$ system
are uncertain at the 0.05--0.10 mag level because the preimaging observations
were not all taken in photometric conditions. This is not a big concern
however, because the method uses relative luminosities. Photometry
of G09 clusters was taken from previous work, with the exception of
NGC6528 (see below).

\subsection{Measuring equivalent widths \label{sub:Measuring-equivalent-widths-1}}

Equivalent widths were measured for the two strongest CaT lines in
the co-added spectra by fitting a model profile over the line central
bandpasses as defined by AD91, and then by computing the encompassed
area. In AD91 Gaussian functions were used, which provide an excellent
fit to CaT line profiles for clusters with metallicities up to 47~Tuc
(${\rm [Fe/H]}\sim-0.7$~dex). On the other hand it was realized
by Rutledge et al. (\cite{Ru97}) and Cole et al. (\cite{cole_etal04},
C04) that CaT lines of stars in more metal-rich clusters have profiles
that are not Gaussian because of the strong damping wings, and they
proposed to fit Moffat functions or sums of Gaussian and Lorentzian
functions. In this work we tried to have an independent view on this
problem, and we used calibration clusters to test both Gaussian and
Gaussian plus Lorentzian (G+L) functions. Gaussian fits were performed
as described in AD91, and we followed C04 for the G+L fits. These
adopted a common line centre $\lambda_{m}$ and the best-fit parameters
were determined using a Levenberg-Marquardt least-squares algorithm
(see Markwardt~\cite{markwardt09}). To estimate the equivalent width
measurement errors, we also measured the \caii\ line strengths independently
on the individual spectra of each star. The calibration clusters of
G09 were already fit with G+L functions, so only Gaussian fits were
performed in this case, to obtain the corresponding $\Sigma W$ values.
The resulting $\Sigma W$ obtained with the two methods are compared
in Appendix~\ref{sec:Transformation-between-the}, where we show
that a linear transformation exists between the two sets of measurements.
We therefore defined G09 as our reference $\Sigma W$ scale, and transformed
into this system the widths computed with the AD91 method.

Having verified that there is a one-to-one correspondence between
widths measured with any of the two methods, we measured metal-rich
program clusters with G+L fits, and for metal-poor program clusters
$\Sigma W$ were computed with Gaussian fits and transformed into
the G09 scale. The metal-poor group contains clusters Pyxis, Rup106,
NGC5824, NGC6139, Ter~3, NGC6325, HP1, NGC6558, NGC6569, M22, M54
and NGC7006, while the metal-rich group contains clusters NGC6356,
NGC6380, NGC6440, NGC6441, and Ter~7. In addition NGC2808, Lynga~7
and Pal~7 were measured with both methods to further confirm the
reliability of the transformation defined in Appendix~\ref{sec:Transformation-between-the}.

Coordinates, radial velocities, $V-V_{{\rm HB}}$ values, and equivalent
widths for the cluster member stars are published in tables to be
found in the electronic version of the paper (see Tables~\ref{tab:Calibration-Cluster-Data}
and \ref{tab:Program-Cluster-Data} for an example of the layout).
These are the fundamental measurements of this work, which allow a
different metallicity calibration to be applied in the future, should
it become available.

\begin{table*}
\caption{Calibration Cluster Data\label{tab:Calibration-Cluster-Data}}

\begin{raggedright} %
\begin{tabular}{lllllllll}
 &  &  &  &  &  &  &  & \tabularnewline
\hline 
\hline 
\multicolumn{1}{c}{ID } & RA (2000)  & Dec (2000)  & Rad Vel  & $\Delta V$  & $W_{8542}$  & $\epsilon$  & $W_{8662}$  & $\epsilon$\tabularnewline
 &  &  & ($\kms$)  & (mag)  & \multicolumn{2}{c}{ (\AA{}) } & \multicolumn{2}{c}{ (\AA{}) }\tabularnewline
\hline 
 &  &  &  &  &  &  &  & \tabularnewline
NGC3201 1\_760  & $154.45472$  & $-46.41605$  & $486$  & $-1.02$  & $2.57$  & $0.14$  & $1.99$  & $0.11$\tabularnewline
NGC3201 1\_2825  & $154.43903$  & $-46.40734$  & $484$  & $-0.95$  & $2.44$  & $0.14$  & $1.86$  & $0.11$\tabularnewline
NGC3201 1\_4251  & $154.43742$  & $-46.40153$  & $487$  & $-2.12$  & $2.69$  & $0.12$  & $2.16$  & $0.10$\tabularnewline
NGC3201 1\_5837  & $154.43194$  & $-46.39489$  & $487$  & $-0.24$  & $2.19$  & $0.21$  & $1.65$  & $0.12$\tabularnewline
NGC3201 1\_6876  & $154.41135$  & $-46.39017$  & $501$  & $-1.60$  & $2.71$  & $0.14$  & $2.07$  & $0.08$\tabularnewline
 &  &  &  &  &  &  &  & \tabularnewline
\hline 
\end{tabular}

\end{raggedright}

\raggedright{}\tablefoot{This table is available in its entirety
in a machine-readable form in the online journal. A portion is shown
here for guidance regarding its form and content. These data supersede
those of Table 1 in Da Costa et al. (2009).} 
\end{table*}

\begin{table*}
\caption{Program Cluster Data \label{tab:Program-Cluster-Data}}

\begin{raggedright} %
\begin{tabular}{llllrllll}
 &  &  &  &  &  &  &  & \tabularnewline
\hline 
\hline 
 & RA (2000)  & Dec (2000)  & Rad Vel  & $\Delta V$  & $W_{8542}$  & $\epsilon$  & $W_{8662}$  & $\epsilon$ \tabularnewline
 &  &  & ($\kms$)  & (mag)  & \multicolumn{2}{c}{ (\AA{}) } & \multicolumn{2}{c}{ (\AA{}) }\tabularnewline
\hline 
 &  &  &  &  &  &  &  & \tabularnewline
Pyxis 1\_1935  & $136.97620$  & $-37.21546$  & $41$  & $-1.11$  & $2.31$  & $0.25$  & $2.15$  & $0.27$\tabularnewline
Pyxis 1\_3319  & $136.97864$  & $-37.20316$  & $50$  & $-0.09$  & $2.52$  & $0.56$  & $2.08$  & $0.72$\tabularnewline
Pyxis 1\_3370  & $137.02170$  & $-37.20279$  & $41$  & $-1.03$  & $2.78$  & $0.34$  & $1.67$  & $0.27$\tabularnewline
Pyxis 1\_4171  & 137.04704  & $-$37.19520  & 45  & 0.02  & \ldots{}  & \ldots{}  & \ldots{}  & \ldots{}\tabularnewline
Pyxis 2\_11646  & 136.98251  & $-$37.27163  & 67  & 0.39  & \ldots{}  & \ldots{}  & \ldots{}  & \ldots{}\tabularnewline
Pyxis 2\_13036  & 136.97849  & $-$37.25888  & 17  & $-$0.16  & \ldots{}  & \ldots{}  & \ldots{}  & \ldots{}\tabularnewline
Pyxis 2\_14488  & $137.01187$  & $-37.24609$  & $58$  & $-0.10$  & $2.27$  & $0.45$  & $1.36$  & $0.52$\tabularnewline
Pyxis 2\_16070  & $136.99621$  & $-37.23422$  & $49$  & $0.00$  & $2.89$  & $0.47$  & $1.86$  & $0.61$\tabularnewline
 &  &  &  &  &  &  &  & \tabularnewline
\hline 
\end{tabular}

\end{raggedright}

\tablefoot{This table is available in its entirety in a machine-readable
form in the online journal. A portion is shown here for guidance regarding
its form and content. The M22 data in this table supersede those of
Table 2 in Da Costa et al. (2009).} 
\end{table*}

\subsection{Reduced equivalent widths \label{sub:Reduced-equivalent-widths-1}}

For calibration clusters, the fixed-slope linear fits in the $\Sigma W$
vs. $V-V_{{\rm HB}}$ plane were performed for $\Sigma W$ values
on both systems (G09 or AD91 converted to G09), and the resulting
$\W$ values were averaged. As described in Da Costa et al. (\cite{dacosta_etal09}),
the fits were computed only for stars with $V-V_{{\rm HB}}$ $\leq$
0.2 because the $(\Sigma W,$ $V-V_{{\rm HB}})$ relation appears
to notably flatten at lower luminosities, as predicted by models (see,
e.g., Carrera et al. \cite{carrera+07}, Starkenburg et al. \cite{starkenburg+10}).
This effect is shown particularly by the NGC6397 and M10 stars in
Fig.~\ref{fig:cal_clusters_ew_fig-1} in which we show the calibration
lines for the entire set of calibration cluster data. The $\Sigma W$
values adopted in the figure are the converted AD91 ones, but in both
cases we found that the best-fitting slope is $a=-0.627$.

This slope was then adopted to compute reduced equivalent widths for
program clusters in both metallicity groups. In the case of NGC2808,
Lynga~7 and Pal~7, which were measured with both methods, the same
procedure as for the calibration clusters was adopted, and their $\W$
values are the average of the two results.

Once reduced equivalent widths $W^{\prime}$ are computed, their average
$\left\langle W^{\prime}\right\rangle $ can be calibrated onto a
metallicity scale using standard clusters, as the next section explains.
Finally, using the calibration relation, the metallicity of each star
for all clusters in our database can be computed by converting its
$W^{\prime}$ to {[}Fe/H{]}, and a representative average metallicity
can also be computed for each cluster.

We note that, particularly in the differentially reddened clusters,
the possibility that AGB stars are included in our ``RGB'' samples
may lead to additional scatter in the ($\Sigma W$, $V-V_{{\rm HB}}$)
plane, and thus to increased uncertainty in the derived abundances.
On the other hand this effect is expected to be small, as shown by
Cole et al. (\cite{cole+00}).

\subsubsection{NGC6553 and NGC6528}

\begin{figure}
\noindent \begin{centering}
\textbf{\includegraphics[width=0.9\columnwidth]{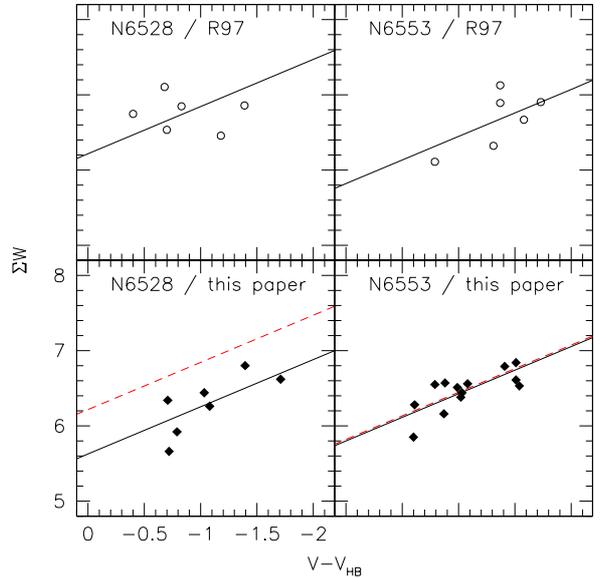}} 
\par\end{centering}

\caption{Comparison of $\Sigma W$ vs. $V-V_{{\rm HB}}$ using both our and
R97 data. The R97 EWs were corrected using the factor 1.117 found
in Sect.~\ref{sub:A-merged-catalog}. The solid lines are linear
fits with fixed slope $a=-0.627$. When our data are plotted, the
fitted lines to R97 data are also shown as dashed lines. A radial
velocity cut was imposed to R97a data to select cluster members: $\Delta$RV$>38\,\kms$
and $\Delta$RV$<52.75\,\kms$ define cluster members for NGC6528
and NGC6553, respectively. \label{fig:Comparison-of-SigmaW} }
\end{figure}

All metallicity determinations in the literature put NGC6553 at a
lower {[}Fe/H{]} than NGC6528, so it is somewhat surprising that in
this work we found NGC6553 at a higher $\W$ than NGC6528. As Table~\ref{tab:Input-data-for-1}
indicates, $\W=5.68\pm0.09$ for NGC6528 and $\W=5.84\pm0.03$ for
NGC6553. Rutledge et al. (\cite{R97a}) found instead $\W=5.41\pm0.14$
and $5.13\pm0.09$ for the two clusters, respectively, which is consistent
with their ranking in {[}Fe/H{]}. To understand the origin of this
discrepancy we show in Fig.~\ref{fig:Comparison-of-SigmaW} the $\Sigma W$
vs. $V-V_{{\rm HB}}$ diagrams for the two clusters, using both R97
data and ours. Evidently, the problem is with NGC6528, for which we
measure EWs that are $\sim$0.6 smaller than those of R97a. It should
be recalled that measuring the strengths of CaT lines in high metallicity
cluster stars is more difficult than at lower metallicities, because
of the stronger presence of metal lines at lower temperatures. These
contaminate line wings, which is where the increased strength resides,
and they also make more difficult to define the appropriate pseudo-continuum.
The veiling caused by molecular species such as TiO and CN is also
a potential concern at lower effective temperatures. The presence
of TiO depresses the pseudo-continuum and reduces the apparent strength
of the CaT lines (see Olszewski et al. \cite{olszewski+91}), and
it might explain our small EWs. However, none of the NGC6528 members
stars observed here or in Gullieuszik et al (\cite{MG09}) show any
evidence for the presence of TiO bands in the part of the spectrum
considered here%
\footnote{In the case of NGC6553, five of the eighteen stars used to calculate
the cluster mean velocity (see Table~\ref{tab:Radial-Velocities})
show definite indications of the presence of the $\lambda$8440~\AA{}TiO
bandhead. Consequently, these five stars have not been used in the
line strength determinations for this cluster. The spectra of the
remaining thirteen NGC6553 stars show no sign of the presence of TiO
bands.%
}. Another possibility is that our $V-V_{{\rm HB}}$ is $\sim$1~mag
larger than that of R97a. There are no stars in common between the
two datasets, however we have five stars in common with van den Bergh
\& Younger (\cite{vdb_younger79}; VY79), which tell us that we are
fainter by $0.07$ mag in the $V$-band. R97a used photometry from
Ortolani et al. (\cite{ortolani+92}), and found that it is fainter
than that of VY79 by $0.05$ mag in the same band. Therefore our zero-points
should agree with those of R97a within $0.02$ mag. In addition R97a
adopted $V_{{\rm HB}}=17.1\pm0.1$ from Ortolani et al. (1992), which
is the same value that we used. Therefore it appears that a zero-point
difference between us and R97a must be ruled out, but it is still
possible that R97a measured six stars among which some had high extinction.
For example excluding the faintest two stars from the R97a data, brings
their average EW closer to our values by $\sim$0.2~\AA{}. Finally
it is of course possible that either our \caii\ line fits or those
of R97a suffer from a still unidentified source of error. In this
respect we can only remark that the trend defined by our data in Fig.~\ref{fig:Comparison-of-SigmaW}
seems more `natural' than that shown by the R97 ones. Clearly an independent
study is needed to solve this puzzle, possibly based on near infrared
photometry like the works of Warren \& Cole (\cite{warren_cole09})
and Lane et al. (\cite{lane+11}).

\subsection{Calibration of reduced equivalent widths}

\begin{table*}
\caption{Input data for the $\left\langle W^{\prime}\right\rangle $ -- {[}Fe/H{]}
relation. \label{tab:Input-data-for-1}}

\begin{centering}
\begin{tabular}{lllr@{\extracolsep{0pt}.}lr@{\extracolsep{0pt}.}lllllll}
 &  &  & \multicolumn{2}{c}{ } & \multicolumn{2}{c}{ } &  &  &  &  &  & \tabularnewline
\hline 
\hline 
NGC  & Alt.  & $\left\langle W^{\prime}\right\rangle $  & \multicolumn{2}{c}{$\epsilon$ } & \multicolumn{2}{c}{{[}Fe/H{]} } & $\epsilon$  &  & {[}Fe/H{]}  & $\epsilon$  & {[}Fe/H{]}  & src\tabularnewline
\hline 
 &  &  & \multicolumn{2}{c}{} & \multicolumn{2}{c}{} &  &  &  &  &  & \tabularnewline
 &  &  & \multicolumn{2}{c}{} & \multicolumn{3}{c}{C09} &  & \multicolumn{2}{c}{CG97} & KI03  & \tabularnewline
\cline{1-3} \cline{6-13} 
 &  &  & \multicolumn{2}{c}{} & \multicolumn{2}{c}{} &  &  &  &  &  & \tabularnewline
 &  &  & \multicolumn{2}{c}{ } & \multicolumn{2}{c}{ } &  &  &  &  &  & \tabularnewline
3201  &  & 3.70  & 0 & 03  & -1 & 51  & 0.02  &  & -1.23  & 0.05  & -1.46  & 3 2 \tabularnewline
4372  &  & 2.33  & 0 & 03  & -2 & 19  & 0.08  &  & ...  & ...  & -2.29  & 1 2 \tabularnewline
4590  & M68  & 1.85  & 0 & 05  & -2 & 27  & 0.04  &  & -1.99  & 0.06  & -2.43  & 1 2 \tabularnewline
5904  & M5  & 4.39  & 0 & 02  & -1 & 33  & 0.02  &  & -1.11  & 0.03  & -1.32  & 1 2 \tabularnewline
5927  &  & 5.29  & 0 & 04  & -0 & 29  & 0.07  &  & ...  & ...  & -0.67  & 3 2 \tabularnewline
6121  & M4  & 4.36  & 0 & 05  & -1 & 18  & 0.02  &  & -1.19  & 0.03  & -1.22  & 3 2 \tabularnewline
6171  & M107  & 4.59  & 0 & 05  & -1 & 03  & 0.02  &  & ...  & ...  & -1.1  & 1 2 \tabularnewline
6254  & M10  & 3.37  & 0 & 02  & -1 & 57  & 0.02  &  & -1.41  & 0.02  & -1.48  & 3 2 \tabularnewline
6397  &  & 2.61  & 0 & 04  & -1 & 99  & 0.02  &  & -1.82  & 0.04  & -2.12  & 1 2 \tabularnewline
6528  &  & 5.68  & 0 & 09  & 0 & 07  & 0.07  &  & ...  & ...  & ...  & 1 3 \tabularnewline
6553  &  & 5.84  & 0 & 03  & -0 & 16  & 0.06  &  & ...  & ...  & ...  & 3 2\tabularnewline
6752  &  & 3.84  & 0 & 02  & -1 & 55  & 0.01  &  & -1.42  & 0.02  & -1.46  & 1 2 \tabularnewline
6838  & M71  & 5.09  & 0 & 04  & -0 & 82  & 0.02  &  & -0.7  & 0.03  & -0.82  & 3 2 \tabularnewline
7078  & M15  & 1.69  & 0 & 04  & -2 & 33  & 0.02  &  & -2.12  & 0.01  & -2.45  & 3 2 \tabularnewline
 &  &  & \multicolumn{2}{c}{} & \multicolumn{2}{c}{} &  &  &  &  &  & \tabularnewline
\hline 
\end{tabular}
\par\end{centering}

{\scriptsize \tablefoot{The numbers in column src indicate the source
of ${\left\langle W^{\prime}\right\rangle }$ measurements: (1) G09
paper, (2) AD91 method, (3) G09 method applied to clusters observed
in this run.}} 
\end{table*}

\begin{figure}
\begin{centering}
\includegraphics[width=0.9\columnwidth]{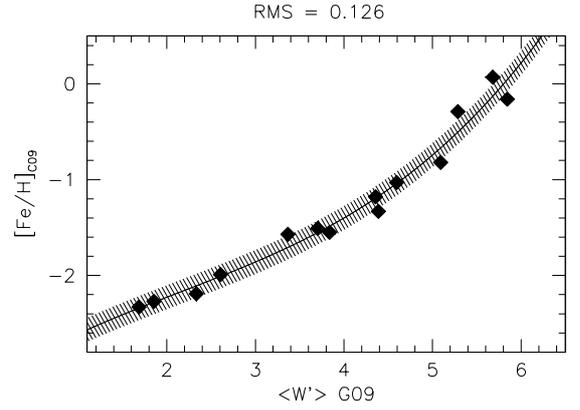} 
\par\end{centering}

\caption{{[}Fe/H{]} on the C09 scale vs. $\left\langle W^{\prime}\right\rangle $
on the G09 scale, from the data in Table~\ref{tab:Input-data-for-1}.
The solid curve shows the cubic fit to the data, and the $\pm{\rm }$RMS
dispersion boundary is represented by the shaded area. Note that the
error bars are smaller than the plot symbols. \label{fig:Same-as-Fig.,-1-1} }
\end{figure}

The data needed to find the transformation between $\W$ and {[}Fe/H{]}
values are summarized in Table~\ref{tab:Input-data-for-1}: for each
calibration cluster we list its $\left\langle W^{\prime}\right\rangle $
and its error, and the metallicity on three different scales. In Carretta
\& Gratton (\cite{CG97}) and Kraft \& Ivans (\cite{KI03}) there
are no clusters more metal rich than ${\rm [Fe/H]}\sim-0.7$ so our
calibration relation is based on the Carretta et al. (\cite{carretta_etal09})
scale, nevertheless values on the other two scales are given for readers
wishing to do comparison studies in a more limited metallicity range.
Using these data we found the following cubic calibration relation
for the C09 scale: 
\[
{\rm [Fe/H]}=0.0178\,\left\langle W^{\prime}\right\rangle ^{3}-0.114\,\W^{2}+0.599\,\W-3.113
\]
 which has a RMS dispersion around the fit of 0.126~dex, and is defined
in the ${\rm [Fe/H]}$ range from $-2.33\,{\rm dex}$ to $+0.07\,{\rm dex}$.
With this relation, $W^{\prime}$ values for each star in each program
cluster can be converted into {[}Fe/H{]} values on the C09 scale,
and average $\W$ can be converted into average metallicities, which
are listed in Table~\ref{tab:metallicities}. The table also lists
the uncertainty in our determinations which is the combination of
the statistical uncertainty in the $\left\langle W^{\prime}\right\rangle $
value, the calibration uncertainty (taken as the RMS dispersion about
the fit) and, where necessary, inclusion of allowance for significant
differential reddening that can affect the individual $V-V_{{\rm HB}}$
values. Also given in the Table are {[}Fe/H{]} values from H10 and
their associated weight (higher numbers indicate more certain values)
as well as the values and their uncertainties from Appendix~1 of
C09.

For the other two metallicity scales, the calibration relations are
${\rm [Fe/H]_{KI03}}=0.496\times\W-3.369$ (RMS=$0.063$ dex) and
${\rm [Fe/H]_{CG97}}=0.383\times\W-2.758$ (RMS=$0.084$~dex). The
range of validity of these two relations is $-2.45$~dex to $-0.67$~dex
for the KI03 scale, and $-2.12$~dex to $-0.7$~dex for the CG97
scale. 
\[
\]

\subsection{Extrapolation of the calibration relation at low $\W$ }

\begin{figure}
\begin{centering}
\includegraphics[width=0.9\columnwidth]{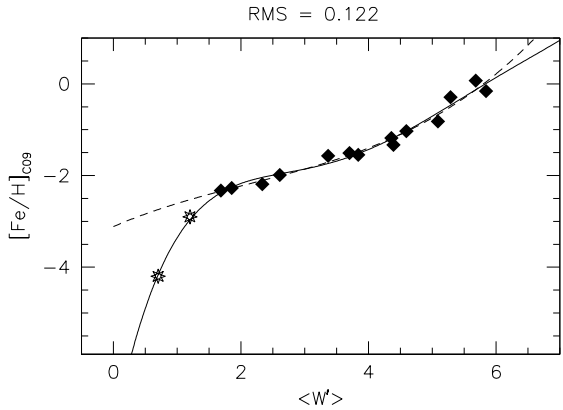} 
\par\end{centering}

\caption{The metallicity - $\W$ relation is plotted here for calibration clusters
(filled diamonds) plus two metal-poor stars from Starkenburg et al.
(\cite{starkenburg+10}, asterisks). Overplotted is a polynomial fit
of degree $n=5$, which shows that our calibration relation (dashed
line) cannot be used outside the metallicity range defined by the
standard clusters. \label{fig:manypol}}
\end{figure}

When metallicity goes to zero, we expect that $\W$ goes to zero as
well, while {[}Fe/H{]} should go to $-\infty$. Therefore our calibration
relation cannot be extrapolated to arbitrarily low $\W$. This is
illustrated by Fig.~\ref{fig:manypol}, where to our calibration
clusters we have added two metal-poor stars from Starkenburg et al.
(\cite{starkenburg+10}). From their Fig.~9 we have estimated the
intercepts of their $(\Sigma(W),V-V_{{\rm HB}})$ relations at $V-V_{{\rm HB}}=0$,
which give $\W$ for a range of metallicities%
\footnote{Note that for single stars an \emph{average} $\W$ does not make sense,
but for simplicity we use that symbol instead of $W^{\prime}$.%
}. In particular we have selected stars CD-38~245 and HD88609, the
two most metal-poor Galactic stars in Starkenburg et al. (\cite{starkenburg+10}).
The two stars have {[}Fe/H{]}$=-4.2$ and $-2.9$, so they can be
used to see the trend of {[}Fe/H{]} vs. $\W$ beyond the range defined
by our calibration clusters. Indeed Fig.~\ref{fig:manypol} shows
a downward trend for $\W\rightarrow0$, as expected. The position
of the two lowest metallicity points is uncertain and their $\W$
are on a slightly different measurement system from that of G09, so
they cannot be used to define a calibration relation valid over a
larger {[}Fe/H{]} range. However for the sake of illustration the
whole data set was fit with a polynomial of degree $n=5$. The figure
makes clear that our fiducial relation cannot be used beyond the {[}Fe/H{]}
range defined by our standard clusters.

\section{Discussion}

\begin{figure*}
\begin{centering}
\includegraphics[width=0.9\columnwidth]{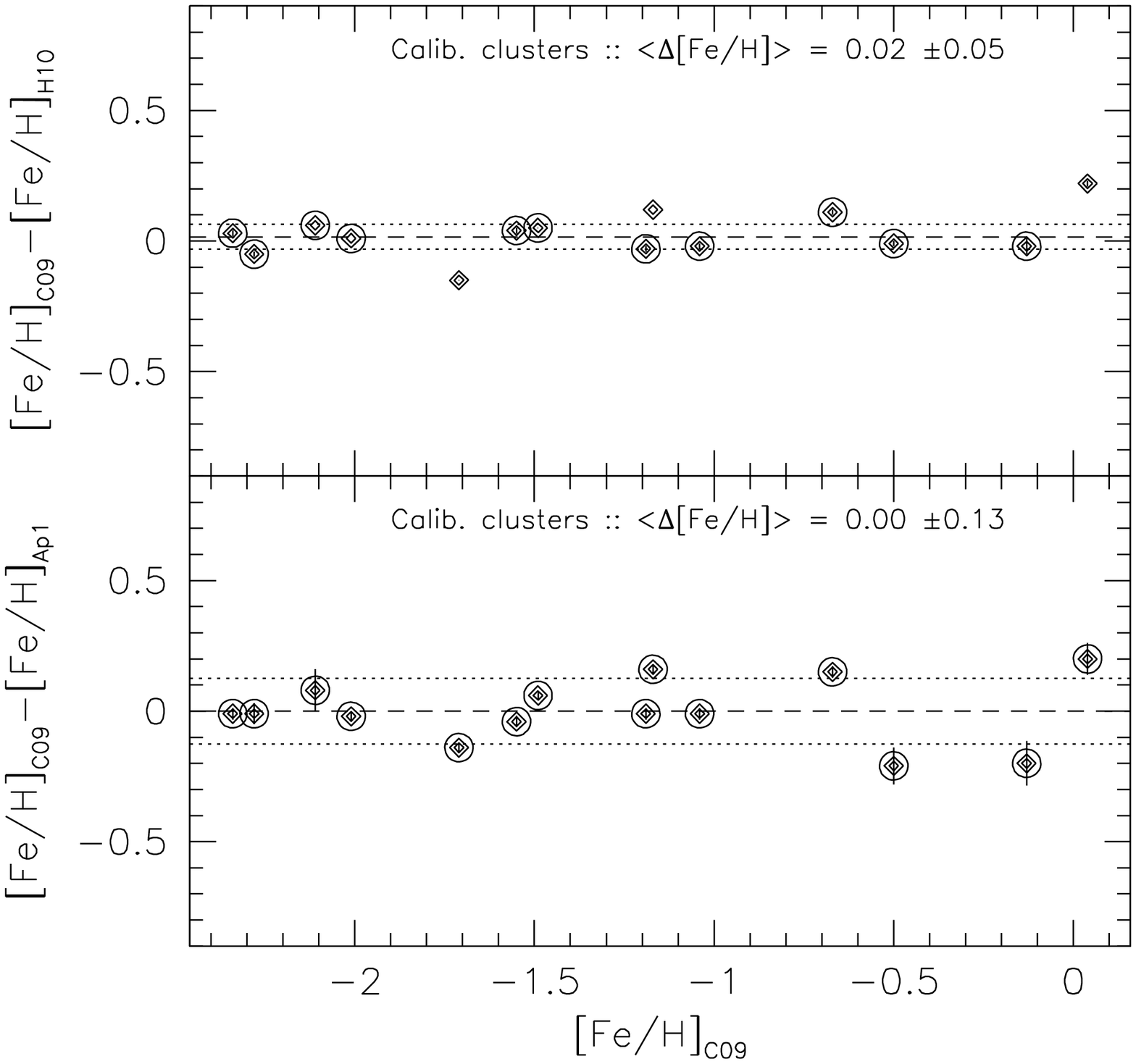}\includegraphics[width=0.9\columnwidth]{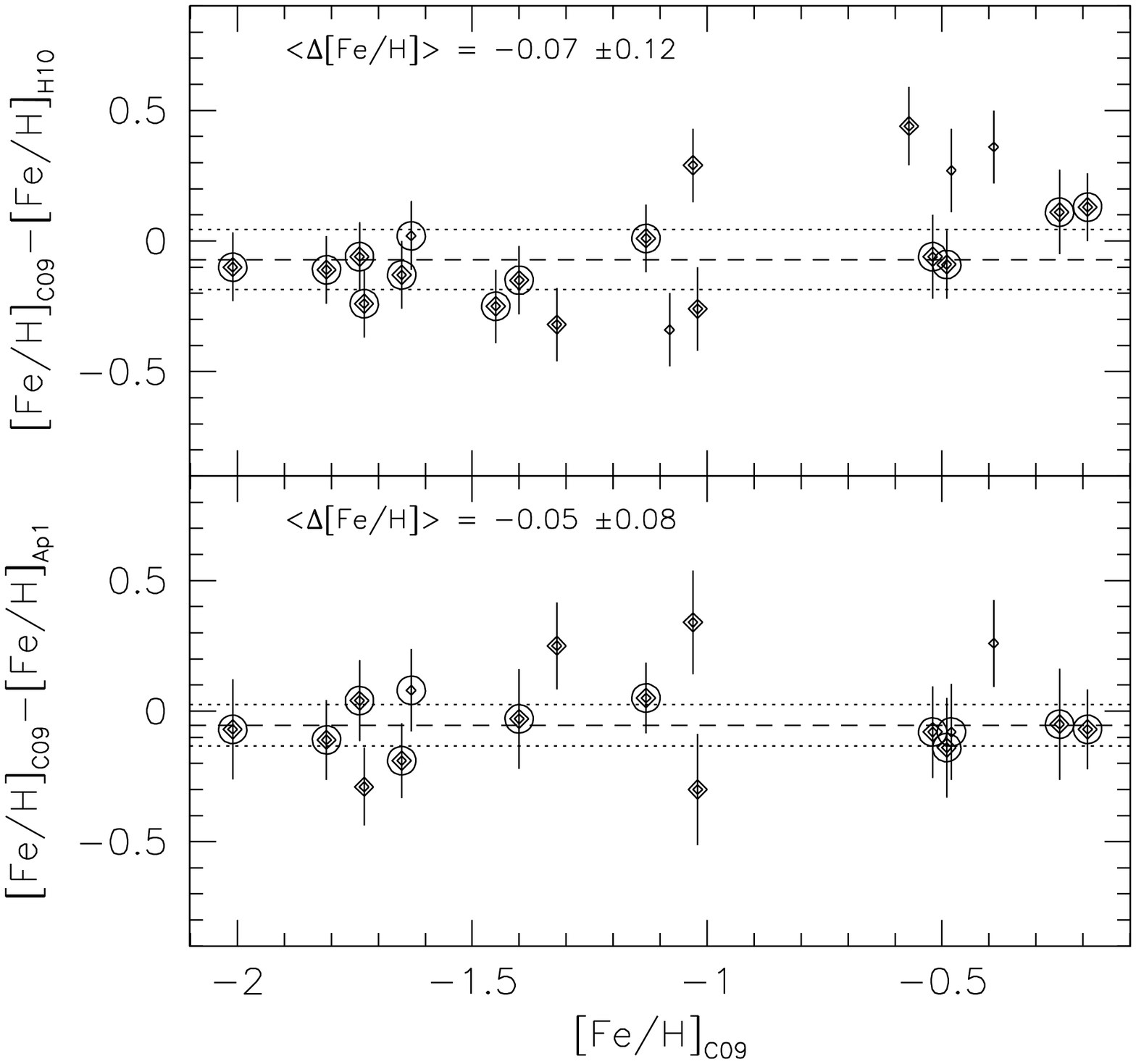} 
\par\end{centering}

\caption{Differences in metallicity between our determinations and those of
the literature are plotted here against {[}Fe/H{]} values computed
in this work. In the upper panels the comparison metallicities are
those of H10, while in the lower panels they are those of Appendix~1
of C09. The dashed and dotted lines represent the average and $\pm\sigma$
of the metallicity differences, which were computed after retaining
clusters with deviations smaller than $1\times\sigma$ from a preliminary
fit (encircled single or double diamonds). The left panels show the
$14$ calibration clusters, and the right panels show the $20$ program
clusters ($17$ for the comparison with C09/Appendix~1, see Table~\ref{tab:metallicities}).
In the right panels single diamonds represent clusters with weight
smaller than $3$ in H10. Error bars in the lower panels represent
the quadratic sum of our errors and those of C09. \label{fig:Differences-in-metallicity}}
\end{figure*}

The average metallicities of the calibration and program clusters
as derived from our CaT measurements are compared to those of H10
and C09 in the panels of Fig.~\ref{fig:Differences-in-metallicity}.
In both cases there is no indication of any systematic deviation with
abundance and the mean offset is close to zero. The agreement indicates
that all three measurements are on a consistent system, which is not
unexpected. The scatter in the abundance differences about the offset
seen in the figure reflects both the uncertainty in our measurements
and in the existing determinations, which generally come from a variety
of heterogeneous sources.

\subsection{Individual Cluster Results}

\begin{flushleft}
\begin{table*}
\caption{Metallicities of program clusters.\label{tab:metallicities}}

\begin{centering}
\begin{tabular}{lllllllllllllll}
 &  &  &  &  &  &  &  &  &  &  &  &  &  & \tabularnewline
\hline 
\hline 
Cluster  & alt.  & $N$  & $\W$  & S  & $\sigma(\W)$  & ${\rm [Fe/H]}$  & $\sigma_{{\rm tot}}$  & ${\rm [Fe/H]}_{{\rm }}$  & w  & ${\rm [Fe/H]}$  & $\epsilon$  & DR?  & $\rho$  & \tabularnewline
 &  &  &  &  &  & C09  &  & H10  &  & ${\rm }_{{\rm C09}}^{{\rm Ap1}}$  &  &  &  & \tabularnewline
\multicolumn{2}{c}{ (1)} & (2)  & (3)  & (4)  & (5)  & (6)  & (7)  & (8)  & (9)  & (10)  & (11)  & (12)  & (13)  & \tabularnewline
\hline 
 &  &  &  &  &  &  &  &  &  &  &  &  &  & \tabularnewline
Pyxis  &  & 5  & 3.90  & 2  & 0.24  & $-$1.45  & 0.14  & $-$1.20  & 3  &  &  &  & 0.8  & \tabularnewline
NGC2808  &  & 17  & 4.46  & 3  & 0.20  & $-$1.13  & 0.13  & $-$1.14  & 4  & $-$1.18  & 0.04  &  & 1.0  & \tabularnewline
Rup106  &  & 9  & 3.29  & 2  & 0.25  & $-$1.74  & 0.13  & $-$1.68  & 6  & $-$1.78  & 0.08  &  & 1.9  & \textifsymbol[ifgeo]{114}\tabularnewline
NGC5824  &  & 17  & 2.60  & 2  & 0.30  & $-$2.01  & 0.13  & $-$1.91  & 4  & $-$1.94  & 0.14  &  & 2.2  & \textifsymbol[ifgeo]{114}\tabularnewline
Lynga7  &  & 8  & 5.21  & 3  & 0.11  & $-$0.57  & 0.15  & $-$1.01  & 3  &  &  & Y  & 0.3  & \tabularnewline
NGC6139  &  & 15  & 3.54  & 2  & 0.15  & $-$1.63  & 0.13  & $-$1.65  & 2  & $-$1.71  & 0.09  &  & 1.0  & \tabularnewline
Ter3  &  & 10  & 4.54  & 2  & 0.12  & $-$1.08  & 0.14  & $-$0.74  & 2  &  &  & Y  & 0.4  & \tabularnewline
NGC6325  &  & 10  & 3.99  & 2  & 0.20  & $-$1.40  & 0.13  & $-$1.25  & 4  & $-$1.37  & 0.14  &  & 1.2  & \textifsymbol[ifgeo]{66}\tabularnewline
NGC6356  &  & 11  & 5.30  & 1  & 0.15  & $-$0.49  & 0.13  & $-$0.40  & 5  & $-$0.35  & 0.14  &  & 0.5  & \tabularnewline
HP1  & BH 229  & 8  & 4.14  & 2  & 0.30  & $-$1.32  & 0.14  & $-$1.00  & 4  & $-$1.57  & 0.09  &  & 1.3  & \textifsymbol[ifgeo]{66}\tabularnewline
NGC6380  & Ton 1  & 8  & 5.31  & 1  & 0.18  & $-$0.48  & 0.16  & $-$0.75  & 2  & $-$0.40  & 0.09  & Y  & 0.7  & \tabularnewline
NGC6440  &  & 8  & 5.56  & 1  & 0.18  & $-$0.25  & 0.16  & $-$0.36  & 6  & $-$0.20  & 0.14  & Y  & 0.6  & \tabularnewline
NGC6441  &  & 7  & 5.27  & 2  & 0.17  & $-$0.52  & 0.16  & $-$0.46  & 6  & $-$0.44  & 0.07  & Y  & 0.6  & \tabularnewline
NGC6558  &  & 4  & 4.61  & 2  & 0.15  & $-$1.03  & 0.14  & $-$1.32  & 5  & $-$1.37  & 0.14  &  & 0.9  & \tabularnewline
Pal7  & IC 1276  & 14  & 5.41  & 3  & 0.12  & $-$0.39  & 0.14  & $-$0.75  & 1  & $-$0.65  & 0.09  & Y  & 0.7  & \tabularnewline
NGC6569  &  & 7  & 4.62  & 2  & 0.19  & $-$1.02  & 0.16  & $-$0.76  & 4  & $-$0.72  & 0.14  & Y  & 0.8  & \tabularnewline
NGC6656  & M22  & 41  & 3.12  & 2  & 0.30  & $-$1.81  & 0.13  & $-$1.70  & 8  & $-$1.70  & 0.08  &  & 2.3  & \textifsymbol[ifgeo]{114}\tabularnewline
NGC6715  & M54  & 15  & 3.30  & 2  & 0.25  & $-$1.73  & 0.13  & $-$1.49  & 7  & $-$1.44  & 0.07  &  & 2.5  & \textifsymbol[ifgeo]{114}\tabularnewline
Ter7  &  & 7  & 5.62  & 1  & 0.12  & $-$0.19  & 0.13  & $-$0.32  & 6  & $-$0.12  & 0.08  &  & 0.4  & \tabularnewline
NGC7006  &  & 18  & 3.49  & 2  & 0.25  & $-$1.65  & 0.13  & $-$1.52  & 6  & $-$1.46  & 0.06  &  & 1.2  & \textifsymbol[ifgeo]{66}\tabularnewline
 &  &  &  &  &  &  &  &  &  &  &  &  &  & \tabularnewline
\hline 
\end{tabular}
\par\end{centering}

{\scriptsize \tablefoot{Column (1) lists the main and alternate
cluster ID. Column (2) is the number of cluster members used in determining
the $\W$ values given in Column (3). Column (4) gives the source
of the $\W$ values: 1 for direct G09 measurements, 2 for transformed
AD91 measurements and 3 for the average of both measurement techniques.
Column (5) is the standard deviation of the $\W$ values. Column (6)
is the {[}Fe/H{]} on the C09 scale which results from applying the
abundance calibration. Column (7) is the total uncertainty in the
{[}Fe/H{]} value: the rms sum of the abundance uncertainties from
the standard error of the mean in $\W$, from the calibration uncertainty,
and from the effects of differential reddening where necessary. Columns
(8) and (9) are the H10 abundance and weight while columns (10) and
(11) are the abundance and error from Appendix 1 of C09 (clusters
with a ``1'' in the Notes column only). Column (12) indicates if
the effects of differential reddening are significant, details are
given in the appropriate sub-section for these clusters. Column (13)
gives the ratio of the r.m.s. dispersion around the linear fit with
fixed slope, to the mean measurement error. In the last column we
mark candidate clusters for metallicity dispersion with a solid arrow
head, while marginal candidates are marked with an open arrow head.
}} 
\end{table*}

\par\end{flushleft}

In the following sections we discuss our new ${\rm [Fe/H]}$ values
for clusters that have issues with (a) data quality, (b) differential
reddening or low statistics affecting the position of the HB, (c)
abundance spread, (d) stars' membership, (e) field contamination,
and (f) large {[}Fe/H{]} differences with the literature. A first
assessment of possible abundance spreads was done by computing, for
each cluster, the ratio $\rho$ of the r.m.s. dispersion around the
linear fit with fixed slope $a=0.627$ \AA~mag$^{-1}$, to the average
measurement error in $\Sigma W$. This parameter is given in the last
column of Table~\ref{tab:metallicities}. Clusters for which $\rho\geq1.5$
are considered candidates for a metallicity dispersion, while clusters
for which $1<\rho<1.5$ are considered marginal candidates. More details
are given in the next sections.

\subsubsection{Pyxis }

The spectra were obtained in the second night, which was partially
cloudy, so the data are of much poorer quality than for the rest of
sample. Three of the 8 stars that were used to determine radial velocity
do not have line strength measures (the Gaussian fit failed). The
given $W'$ value is the weighted mean value, and the HB luminosity
was determined by considering stars within $2'$ of center, and taking
the mean of red HB stars.

Our value of $-1.45\pm0.14$~dex is somewhat lower than that ($-1.20$)
tabulated by H10. It is, however, consistent with the spectroscopic
determination of Palma et al. (\cite{palma+00}) who measured ${\rm [Fe/H]_{ZW}}=-1.4\pm0.1$
from a spectrum at the CaT of a single Pyxis red giant. The CMD based
photometric determinations of Irwin et al. (\cite{irwin+95}), $-1.1\pm0.3$,
and Sarajedini \& Geisler (\cite{sarajedini_geisler96}), $-1.2\pm0.15$
are also not inconsistent with this determination. With its dominant
red horizontal branch morphology Pyxis is clearly a `young halo' object
(cf. Da Costa \cite{dacosta95}, Sarajedini \& Geisler \cite{sarajedini_geisler96}).

\subsubsection{Ruprecht 106}

At face value the dispersion in the $\Sigma W$ values for the 9 candidate
members of this cluster (0.30~\AA{}) is significantly larger than
the mean measurement errors (0.15~\AA{}), suggesting the possible
presence of an intrinsic abundance range in this cluster. On further
investigation, however, such a result seems unlikely. The large apparent
dispersion is driven by two stars: star 2\_17214 which lies below
the fitted $-$0.627~\AA{}~mag$^{-1}$ line and star 2\_14496 which
lies above it. If these two stars are excluded the remaining 7 have
dispersion in $\Sigma W$ of only 0.089~\AA{}and a mean measurement
error of 0.13~\AA{}. Star 2\_17214 is the brightest in the observed
sample and in the color-magnitude diagram (CMD) derived from our pre-imaging
photometry it lies bluer and brighter than the red giant branch. This
CMD location is confirmed when the stars observed spectroscopically
here are cross-identified with the photometric study of Sarajedini
\& Layden (\cite{sarajedini_layden}). Our star 2\_17214 is SL37 and
again it lies significantly to the blue of the well defined red giant
branch in the Sarajedini \& Layden color-magnitude diagram. It is
likely that 2\_17214/SL37 is either an AGB or post-AGB star that is
effectively \textquotedbl{}too bright\textquotedbl{} in $V-V_{{\rm HB}}$
for its $\Sigma W$ value: RGB stars with the same $V-I$ color are
$\sim$0.7 mag fainter and such a magnitude offset would place the
star very close to the fitted line in the $\Sigma W$, $V-V_{{\rm HB}}$
plane.

The second discrepant star, 2\_14496 (SL519), is the faintest in our
sample. With $V-V_{{\rm HB}}$=0.18, it is close to the $V-V_{{\rm HB}}=$+0.2
cutoff for inclusion in the fitting process, for fainter magnitudes
the line strengths do not decrease at the same rate as for more luminous
stars. Using the Sarajedini \& Layden (\cite{sarajedini_layden})
photometry this star has $V-V_{{\rm HB}}=$0.25 and thus would be
automatically excluded from the fit. Its line strength is not inconsistent
with those of the slightly more luminous stars (cf. the fainter M10
and NGC6397 stars in Fig.~\ref{fig:cal_clusters_ew_fig-1}). We conclude
there is no compelling reason to consider this star discrepant. Further,
the other stars in our spectroscopic sample have locations in the
Sarajedini \& Layden (\cite{sarajedini_layden}) CMD consistent with
our photometry and with locations on the RGB. Taken together these
arguments suggest that there is no evidence for any significant abundance
spread in Ruprecht 106. This result is consistent with the earlier
work of Da Costa, Armandroff \& Norris (\cite{dacosta+92}) who observed
7 red giants in this cluster at the \caii\ triplet%
\footnote{There is one star in common with our work: star 2205 of Da Costa et
al. (1992) is star 1\_6406 here.%
}. They found an abundance of {[}Fe/H{]}$=-$1.69$\pm$0.05~dex fully
consistent with that derived here. Moreover, the 7 stars observed
by Da Costa et al. (\cite{dacosta+92}) have a dispersion in $\Sigma W$
about their adopted ($\Sigma W$, $V-V_{{\rm HB}}$) relation of 0.16~\AA{}with
a mean measurement error of 0.25~\AA{}. Again this is consistent
with a lack of any intrinsic abundance dispersion in this cluster.

\subsubsection{An abundance spread in NGC5824 \label{sub:An-abundance-spread/5824}}

\noindent \begin{flushleft}
\begin{figure}
\noindent \begin{centering}
\includegraphics[angle=-90,width=0.9\columnwidth]{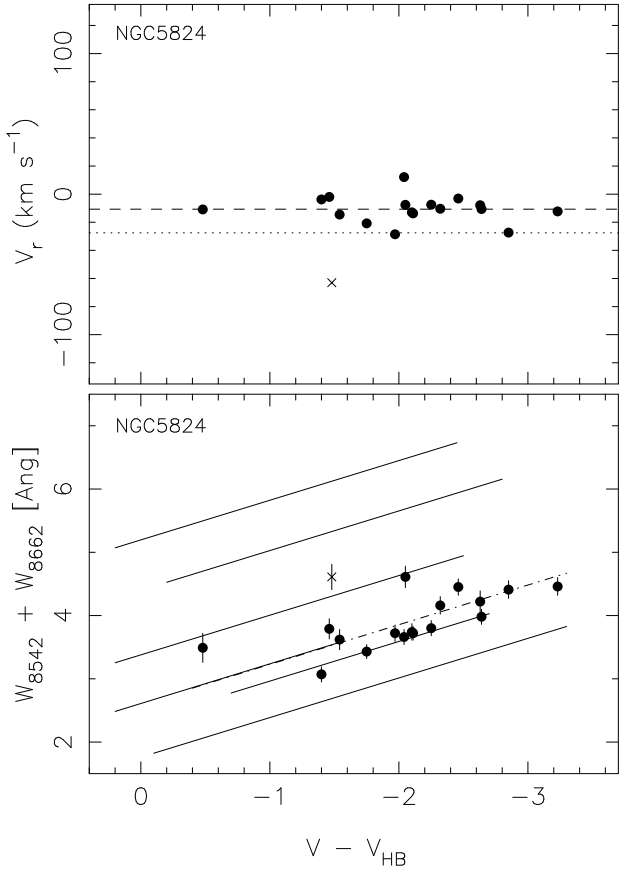} 
\par\end{centering}

\caption{Data for stars observed in NGC5824. The upper panel shows the observed
heliocentric velocity plotted against $V-V_{{\rm HB}}$. The dotted
line is the cluster velocity given by the Harris catalogue, the dashed
line is the mean velocity for the 17 stars selected as cluster members.
The sole star identified as a radial velocity non-member is plotted
as an x-sign. The lower panel shows $\Sigma W$ against $V-V_{{\rm HB}}$
for the same stars. The dot-dash line is the fit of the calibration
relation to the cluster member points (filled circles). Shown also
are the calibration lines for the standard clusters M15, NGC4372,
NGC6397, M10, M5 and M71 (in order of increasing $\Sigma W$). The
significant scatter in the NGC5824 member data suggests the presence
of an intrinsic metallicity dispersion. \textbf{\label{fig:The-metallicity-dispersion-5824}}}
\end{figure}

\par\end{flushleft}

This object is a little studied, luminous ($M_{V}=-8.85$) globular
cluster that lies $32$~kpc from the Sun and $26$~kpc from the
Galactic center (H10). After NGC~2419 for which $M_{V}=-9.4$ and
$R_{{\rm G}}=90$ kpc, NGC5824 is the next most luminous distant globular
cluster in the outer halo -- M54, the nuclear star cluster of the
Sgr dwarf galaxy has ($M_{V}$, $R_{{\rm G}}$) of ($-9.98$, $18.8$)
while NGC~5024 has ($-8.71$, $18.4$) using the data tabulated by
H10. The only previous spectroscopic abundance determination for this
cluster is based on an integrated spectrum at the CaT obtained by
Armandroff \& Zinn (\cite{AZ88}). This is the source of the abundances
of {[}Fe/H{]}$=-1.91$ (weight $4$) and {[}Fe/H{]}$=-1.94\pm0.14$
listed by H10 and C09, respectively.

In the lower panel of Fig.~\ref{fig:The-metallicity-dispersion-5824}
we show our $\Sigma W$ measures plotted against $V-V_{{\rm HB}}$
for the $17$ members observed in this cluster, together with the
calibration lines for the standard clusters. The measurements were
made on the AD91 system but have been transformed to the G09 system.
The value of $\W$ for these stars leads to an abundance of {[}Fe/H{]}$=-2.01\pm0.13$
for NGC5824, which is in excellent agreement with the earlier determination.
However, while the effect is not as striking as it is for M22 (see
Da Costa et al 2009 Fig. 4), it does appear in Fig.~\ref{fig:The-metallicity-dispersion-5824}
that the dispersion of the NGC5824 $\Sigma W$ values is larger than
would be expected from the measurement errors alone. Specifically,
the standard deviation about the fitted line is $0.31$~\AA{}\ while
the mean measurement error in $\Sigma W$ is $0.14$~\AA{}, \emph{suggesting
the presence of an internal abundance spread in NGC5824}. The largest
contribution to the measured dispersion in the $\Sigma W$ values
comes from the stars 2\_32429 and 1\_8575, which have $\Sigma W$
values 0.72 and 0.59~\AA{}, respectively, higher than that expected
from the mean relation at their $V-V_{{\rm HB}}$ values. These two
stars are not distinguished from the other NGC5824 stars observed
in terms of radial velocity (see upper panel of Fig.~\ref{fig:The-metallicity-dispersion-5824}),
distance from the cluster center, or location in the color-magnitude
diagram. Further, the $\Sigma W$ measurement error for star 2\_32429,
0.17~\AA{}, is consistent with the mean for the other stars. The
value for 1\_8575 is somewhat higher at 0.23~\AA{}, but that is not
surprising given it is the faintest star in the observed sample. If
this star is excluded the dispersion in the $\Sigma W$ values is
reduced only marginally from 0.31~\AA{}to 0.28~\AA{}, while the
mean error in $\Sigma W$ is unchanged at 0.14~\AA{}. We conclude
that despite the small number of stars observed there is strong evidence
for intrinsic line strength variations in our NGC5824 sample. For
comparison, the equivalent numbers for M22 are ($0.30$, $0.11$)
with a sample of $41$ stars, while for M54, a cluster also known
to have an internal abundance spread, e.g. Carretta et al. (\cite{carretta_etal10},
hereafter Car10), they are ($0.25$, $0.16$) for a sample of $15$
stars (see below). These are the only three clusters in our entire
sample where there is substantial evidence for the presence of an
intrinsic {[}Fe/H{]} dispersion. Thus NGC5824 now joins the small
number of globular clusters where such intrinsic {[}Fe/H{]} dispersions
are known.%
\footnote{From a preliminary analysis of high resolution spectra of three stars
from our sample Villanova \& Geisler (in preparation) find a significant
{[}Fe/H{]} difference, much larger than the errors.%
}

If we subtract the contribution from the equivalent width measurement
error and apply the abundance calibration, then the corresponding
intrinsic abundance dispersion in NGC5824 is $\sigma_{{\rm int}}({\rm [Fe/H]})=0.12$~dex,
which is comparable to our results for M22 ($\sigma_{{\rm int}}({\rm [Fe/H]})=0.15$
dex, Da Costa et al. 2009). A dispersion of this order places NGC5824
at a consistent location for its luminosity in Fig.~7 of Car10. With
a sample of only $17$ stars the form of the abundance distribution
is not well constrained but the impression from Fig.~\ref{fig:The-metallicity-dispersion-5824}
is that it may be similar to those of M22 (Da Costa et al., 2009)
and $\omega$Cen (e.g., Johnson \& Pilachowski \cite{johnson_pilachowski10})
with a steep rise on the metal-poor side to a peak at {[}Fe/H{]}$=-2.06$
and a broader tail to {[}Fe/H{]}$\approx-1.7$~dex. The median abundance
is {[}Fe/H{]}$=-2.02$~dex. We have accepted programs at Gemini-South
with GMOS and at the VLT with FORS2 to substantially increase the
number of NGC5824 red giants with abundance determinations.

The discovery of a probable intrinsic abundance spread in NGC 5824
is particularly intriguing as Newberg et al. (\cite{newberg+09})
have recently suggested that this cluster is possibly the former nucleus
of a dwarf galaxy whose tidal disruption is responsible for the stellar
stream known as the Cetus Polar Stream (Newberg et al. \cite{newberg+09}).
The Cetus Polar Stream is a low metallicity ({[}Fe/H{]}$\approx-2.1$)
tidal stream approximately $34$~kpc from the Sun. A connection between
the stellar stream and NGC5824 would strengthen the hypothesis that
there exists a set of globular cluster-like stellar systems, characterized
by the presence of internal {[}Fe/H{]} abundance ranges and higher
than average luminosity, which have their origin as former dwarf galaxy
nuclei (e.g., $\omega$Cen) or dwarf galaxy central star clusters
(e.g., M54) and which are distinct from ``regular'' globular clusters
(e.g. Da Costa et al. 2009, Car10).

\subsubsection{Lynga 7 }

Stars within a radius of $1\arcmin$ from the cluster center were
initially selected from the photometry dataset to define the $V_{{\rm HB}}$
value and to define the cluster locus in the $(\Sigma W,V-V_{{\rm HB}})$
plane. This process and the individual radial velocities indicated
$5$ stars as probable members. A further $4$ stars were then added
to the sample based on radial velocities and line strengths consistent
with the probable members. However, in calculating the value of $\left\langle W^{\prime}\right\rangle $,
the brightest star (1\_4175) was excluded as its spectrum clearly
reveals the presence of TiO absorption bands that affect the strength
of the CaT features. We note also that our CMD clearly shows the effects
of differential reddening across the cluster in that the blue end
of the horizontal branch is $\sim0.5$~mag brighter than the red
end. We used the midpoint value for $V_{{\rm HB}}$ and note that
a $\pm0.13$~mag uncertainty in $V_{{\rm HB}}$ ($\pm$ one quarter
of the full range) gives an additional uncertainty of $\pm0.06$~dex
in the abundance determination. Our determination of ${\rm [Fe/H]_{C09}}=-0.57\pm0.15$
is larger than that tabulated by H10 but is entirely consistent with
the previous spectroscopic determination of Tavarez \& Friel (\cite{tavarez_friel95})
who found ${\rm [Fe/H]}=-0.62\pm0.15$ from an analysis of Fe line
strengths in moderate resolution blue spectra of $4$ Lynga~7 red
giants. The relatively high abundance is consistent with the interpretation
of Lynga~7 as a (thick) disk globular cluster (cf. Ortolani et al.
\cite{ortolani+93}).

\subsubsection{NGC6139}

Our {[}Fe/H{]} value is the first spectroscopic determination for
this cluster, previous determinations derive from the integrated light
$Q_{39}$ photometry index of Zinn (\cite{zinn80}), or from the colour
of the red giant branch in the CMD.

\subsubsection{Terzan 3 }

Three probable members included in the calculation of the cluster
radial velocity were excluded from the $W'$ determination as they
are clearly fainter than $V-V_{{\rm HB}}=+0.2$. As for Lynga~7,
this cluster suffers from notable differential reddening with the
blue end of the predominantly red horizontal branch being approximately
$0.6$~mag brighter than the red end. The midpoint was used as the
$V-V_{{\rm HB}}$ value. Again assuming the uncertainty in this value
is $\pm$ one quarter of the full range, the corresponding uncertainty
in the derived abundance from the effects of differential reddening
is $0.06$~dex. The only previous spectroscopic determination of
the abundance of Terzan~3 is that of Cote (\cite{cote99}), who found
{[}Fe/H{]}$=-0.75\pm0.25$ from an analysis of the strengths of strong
Fe~I lines on low S/N high resolution spectra, which is $0.33$~dex
higher than our value. Given the lack of well determined abundances
for the clusters used to calibrate his line strength - abundance relation,
the discrepancy with our determination is not significant.

\subsubsection{NGC6325}

This cluster is marked in Table~\ref{tab:metallicities} as having
a marginal metallicity spread. The dispersion of $\Sigma W$ around
the fit vs. $V-V_{{\rm HB}}$ is inflated by one star only, 2\_62566,
which is also the faintest in the sample. If that star is removed
from the fit, then the ratio of the rms dispersion around the fit
to the average error in $\Sigma W$ becomes $\rho=0.92.$ We conclude
that there is no evidence of a metallicity dispersion in NGC6325.

\subsubsection{HP1 }

This cluster is projected against a dense Galactic bulge field and
consequently isolating candidate members is not straightforward. We
began by considering the CMD for stars within $23\arcsec$ of the
cluster center (cf. Ortolani et al. \cite{ortolani+97}) to identify
the blue horizontal branch of the cluster and thus establish $V_{{\rm HB}}$.
The resulting $(\Sigma W,V-V_{{\rm HB}})$ and radial velocity data
then allowed the selection of $6$ probable members within $1\arcmin$
of the cluster center and a further $2$ probable members at larger
radial distances. Of the 6 more central stars, 5 are confirmed as
members also by the proper motion analysis of Ortolani et al. (\cite{ortolani+11}),
while the other 3 are outside their explored area. The abundance determination
is based on all $8$ candidate members but is unchanged if only the
six inner stars are considered.

Our abundance {[}Fe/H{]}$=-1.32\pm0.14$ is somewhat lower than the
value $-1.0\pm0.2$ given by Barbuy et al. (\cite{barbuy+06}) which
was based on high dispersion spectra of two stars. Since {[}Ca/Fe{]}=0.03
was obtained in Barbuy et al. (\cite{barbuy+06}), the lower {[}Fe/H{]}
value found here could not be explained by a Ca overabundance, as
it may be the case for other clusters. The stars analyzed by Barbuy
et al. (\cite{barbuy+06}), HP1-2 and HP1-3, correspond to our stars
1\_6931 and 1\_4996 and these are among our probable cluster members.
Our lower abundance alleviates to some extent the disparity between
the blue horizontal branch morphology of this cluster and the relatively
high abundance found by Barbuy et al. (\cite{barbuy+06}). We note
also that recently Valenti et al (\cite{valenti+10}) determined an
abundance of {[}Fe/H{]}$_{{\rm CG97}}=-1.12\pm0.2$~dex for HP1 based
on the infrared colors of the red giant branch stars.

In Table~\ref{tab:metallicities} the value of $\rho$ derived from
the 8 candidate members suggests the possible existence of an intrinsic
abundance spread in this cluster. However, we do not think this is
likely to be the case. Rather we suggest that a combination of differential
reddening, increased photometric errors due to the crowded nature
of the field, and the possible inclusion of AGB stars in the candidate
member sample, have all contributed to the scatter about the fitted
line marginally exceeding that expected from the measurement errors.
Certainly the CMD for stars within 1$\arcmin$ of the cluster center
shows considerable scatter, and without a larger sample of confirmed
members it is difficult to convincingly identify the cluster RGB and
possible AGB sequences. Spectroscopic observations of a larger sample
of cluster members would of course also lead to stronger constraints
on the presence or absence of an intrinsic abundance spread in this
cluster.

\subsubsection{NGC6380}

The brightest cluster member (1\_3509) was not included in the calculation
of $\left\langle W^{\prime}\right\rangle $ as the spectrum shows
clear signs of TiO absorption affecting the location of the pseudo-continuum
in the vicinity of the CaT lines. This cluster is also affected by
significant differential reddening with the $V$ magnitude difference
between the blue and red ends of the predominantly red horizontal
branch differing by $\sim0.7$~mag. We again adopted the midpoint
for $V_{{\rm HB}}$. As for Lynga~7 and Terzan~3 above, if we take
$\pm0.18$~mag as the uncertainty in the $V_{{\rm HB}}$ value (i.e.
$\pm$ one quarter of the full range), then there is an additional
uncertainty of $\pm0.09$~dex in our abundance determination.

\subsubsection{NGC6440 }

This cluster is another that suffers from notable differential reddening.
Using the same approach as for the other clusters with differential
reddening, the estimated uncertainty in the adopted $V_{{\rm HB}}$
magnitude is $\pm0.14$~mag which leads to an additional uncertainty
in our {[}Fe/H{]} value of $\pm0.07$~dex.

\subsubsection{NGC6441}

As for NGC6440 this cluster also suffers from differential reddening.
Using the same approach as for the other differentially reddened clusters
the uncertainty in the adopted $V_{{\rm HB}}$ value is $\pm0.15$~mag.
This leads to an additional uncertainty of $\pm0.07$~dex in the
{[}Fe/H{]} value which is included in the total uncertainty listed
in Table~\ref{tab:metallicities}.

\subsubsection{NGC6558 }

The brightest cluster member (1\_4901) in our sample was excluded
from the calculation of $\left\langle W^{\prime}\right\rangle $ as
the spectrum shows clear TiO absorption. Our abundance, {[}Fe/H{]}$=-1.03\pm0.14$,
is in excellent agreement with that, $-0.97\pm0.15$, found by Barbuy
et al. (\cite{barbuy+07}) from an analysis of VLT/FLAMES spectra
of $5$ red giant members. Since {[}Ca/Fe{]}=+0.05 was derived, in
this case the metallicity {[}Fe/H{]}$\approx${[}Ca/H{]}. In contrast,
H10 and C09 list a lower abundance for this cluster. The origin of
those values lies with the integrated photometry and integrated spectroscopy
$Q39$ values discussed in Zinn \& West (\cite{ZW84}). Given our
more reliable determination and the agreement with the results of
Barbuy et al. (\cite{barbuy+07}) our abundance is preferable. The
presence of TiO in the spectrum of the brightest red giant in our
sample is also indicative of a relatively high abundance. We note
that our stars 1\_3494, 1\_4901 and 1\_5672 correspond to Barbuy et
al. (\cite{barbuy+07}) stars F97, F42 and B73. With this relatively
high abundance and blue horizontal branch morphology (Rich et al.
\cite{rich+98}, Barbuy et al. \cite{barbuy+07}, Barbuy et al. \cite{barbuy+09})
the cluster is similar to HP1. As discussed in Barbuy et al. (\cite{barbuy+07}),
these moderately metal-rich globular clusters with blue horizontal
branch morphologies may have been amongst the first objects to form
in the Galactic bulge.

\subsubsection{Pal7}

This cluster is also among those affected by differential reddening.
We have taken the uncertainty in the $V_{{\rm HB}}$ value for the
dominant red horizontal branch in this cluster as $\pm0.16$~mag,
which leads to an additional uncertainty of $\pm0.08$~dex in our
abundance determination. The only other abundance estimate for this
cluster is that of Côté (\cite{cote99}) who derived an abundance
of {[}Fe/H{]}$=-0.75\pm0.25$ from a similar analysis process to that
described in the results for Terzan~3. As noted in the discussion
for that cluster, given the lack of well determined abundances for
the clusters used by Côté (\cite{cote99}) to calibrate his line strength
relation, we consider our determination significantly more reliable.

\subsubsection{NGC6569}

Like many of the clusters in our sample, NGC6569 is also affected
by differential reddening. The uncertainty in the $V_{{\rm HB}}$
value of this red horizontal branch cluster is $\pm0.20$~mag which
leads to an additional uncertainty in our abundance determination
of $\pm0.09$~dex. Our determination is somewhat less than the values
listed by H10 and C09. These have their origin in the $Q39$ value
given by Zinn \& West (\cite{ZW84}), which was derived from integrated
photometry and integrated spectroscopy of the cluster. Recently, Valenti
et al. (\cite{valenti+11}) list {[}Fe/H{]}$=-0.79\pm0.02$ (internal
error only) for this cluster based on high resolution near-IR spectra
of $6$ cluster red giants. This value is only slightly more than
1~sigma different from our determination. There are no stars in common
between our sample and that of Valenti et al. (\cite{valenti+11}).

\subsubsection{M22}

Our results for this cluster have been discussed in Da~Costa et al.
(\cite{dacosta_etal09}). Here we note only that a different mean
abundance results from the use of the C09 abundance scale rather than
that of KI03.

\noindent \begin{flushleft}
\begin{figure}
\noindent \begin{centering}
\includegraphics[angle=-90,width=0.9\columnwidth]{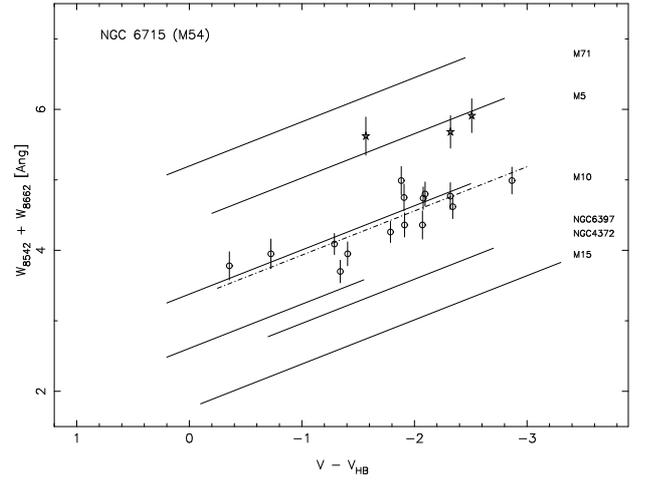} 
\par\end{centering}

\caption{$\Sigma W$ against \vvhb\ for M54 stars compared to calibration
clusters, showing a clear metallicity dispersion. Five-point asterisks
identify the three metal-rich stars that are likely members of the
Sagittarius dwarf galaxy population. \label{fig:The-metallicity-dispersion-m54}}
\end{figure}

\par\end{flushleft}

\subsubsection{M54}

M54 is a luminous globular cluster superposed on the nucleus of the
Sgr dwarf galaxy (e.g. Bellazzini et al. \cite{B+08}). The most recent
detailed study of this system is that of Car10 who analyzed high dispersion
VLT/FLAMES spectra for somewhat more than $100$ stars in the central
field of Sgr. They associate the bulk of their stars with M54 finding
a mean abundance of $-1.56\pm0.02$%
\footnote{Note that Carretta et al. (\cite{carretta_etal09}) listed {[}Fe/H{]}$=-1.44\pm0.07$
for M54.%
} and an intrinsic abundance dispersion of $0.19$~dex ($76$ stars).
The remainder of their stars are relatively metal-rich ({[}Fe/H{]}$\gtrsim-1.2$,
see Fig.~4 of Car10) and Car10 associate them with the Sgr nucleus
population finding a mean abundance of $-0.62\pm0.07$~dex with a
large intrinsic dispersion of $0.35$~dex ($27$ stars).

In Fig.~\ref{fig:The-metallicity-dispersion-m54} we plot our $\Sigma W$
measures against \vvhb\ for the stars observed in our M54 field.
The measurements were made on the AD91 system but have been transformed
to the G09 system. These data show $3$ stars (1\_5389, 2\_20946 and
2\_22704) with relatively large values of $\Sigma W$ while the remaining
$15$ have weaker line strengths. We suggest that these $3$ stars,
which have abundances of {[}Fe/H{]}$=-1.24$, $-0.97$ and $-1.21$,
respectively, belong to the Sgr population rather than M54. We note
that it is not surprising that we did not find any more metal-rich
members of the Sgr population as the original target selection was
deliberately biased towards the M54 RGB population (see Car10 Fig.~3).
The remaining $15$ stars, assumed to be members of M54, have a mean
abundance of $-1.73\pm0.13$~dex. This is in reasonable agreement
with the mean abundance $-1.56\pm0.02$ given in Car10.

As was found for M22 (Da Costa et al. 2009), and for NGC5824 above,
the dispersion in the $\Sigma W$ values in Fig.~\ref{fig:The-metallicity-dispersion-m54}
is notably larger than that expected on the basis of the uncertainties
in the $\Sigma W$ values. The dispersion about the fitted line in
Fig.~\ref{fig:The-metallicity-dispersion-m54} is $0.25$~\AA{}\ while
the mean uncertainty in the $\Sigma W$ values is $0.16$~\AA{}.
Subtracting this error contribution in quadrature then gives an intrinsic
dispersion in $\Sigma W$ of $0.19$~\AA{}, which translates to an
intrinsic abundance dispersion $\sigma_{{\rm INT}}({\rm [Fe/H]})$
of $0.09$~dex. This is smaller than the $0.19$~dex intrinsic dispersion
found by Car10. We have no straightforward explanation for this discrepancy
other than to note that the Car10 sample is five times larger and
may therefore more fully probe the extremes of the underlying distribution.
Uncertainties in the methods used to fix $T_{{\rm eff}}$ might explain
shifts in the absolute metallicities, but they should not affect relative
abundances, and thus abundance dispersions. 

There are only $3$ stars in common between our study and that of
Car10. Our M54 stars 1\_1982, 2\_10700 and 2\_17683 correspond to
Car10 M54 stars 38004437, 38004707 and 38009987, respectively, while
none of our three postulated Sgr stars are in the Car10 list. For
the three stars in common, the differences in the {[}Fe/H{]} values,
in the sense (our value -- Car10) are $0.15$, $-0.21$ and $-0.08$~dex
for a mean difference of $-0.05$~dex with a sigma of $0.18$~dex.
The mean difference is in the same sense as the difference in the
mean abundances ($-0.10$~dex). The dispersion, however, is somewhat
larger than expected given that our individual abundance determinations
have an uncertainty of order $0.11$~dex and those of Car10 nominally
of order $0.02$~dex. Without a larger sample of common objects this
question cannot be investigated further.

\subsubsection{Terzan 7 }

Five probable cluster members were not included in the calculation
of $\W$ as they are fainter than \vvhb$=+0.2$. Their line strengths
are, however, consistent with those of the more luminous members.
Further, two additional probable members were excluded from the $\W$
calculation as their line strength measures were particularly uncertain.

\subsubsection{NGC7006}

Two probable cluster members were excluded from the $\W$ calculation
as they are fainter than $V-V_{{\rm HB}}=+0.2$. Their line strengths
are nevertheless consistent with those of the more luminous members.
For the remaining 18 stars the dispersion in $\Sigma W$ about the
fitted line of slope $-$0.627~\AA{}~mag$^{-1}$ is 0.26 while the
mean measurement error is 0.21~\AA{}. While this might be considered
as evidence for the presence of a small intrinsic metallicity spread,
we do not believe this to be the case. The increased dispersion in
$\Sigma W$ is driven by 3 stars that lie below the fitted line. Inspection
of the location of these nominally discrepant stars in the CMD derived
from our pre-imaging photometry indicates that all three stars are
likely AGB stars. At the same color AGB stars are brighter in $V-V_{{\rm HB}}$
than RGB stars and this magnitude offset can reach 0.5~mag or more
for stars on the lower part of the AGB, as is the case here. It results
in AGB stars lying to the right of RGB stars of similar color in the
($\Sigma W$, $V-V_{{\rm HB}}$) plane mimicking a lower abundance.
If these three AGB stars are excluded from the fit, the dispersion
is reduced to 0.20~\AA{} and equals the mean measurement error. We
conclude that there is no evidence in our data for an intrinsic abundance
spread in this cluster.

\subsection{A merged catalog of globular cluster metallicities \label{sub:A-merged-catalog}}

\begin{figure}
\begin{centering}
\includegraphics[width=0.9\columnwidth]{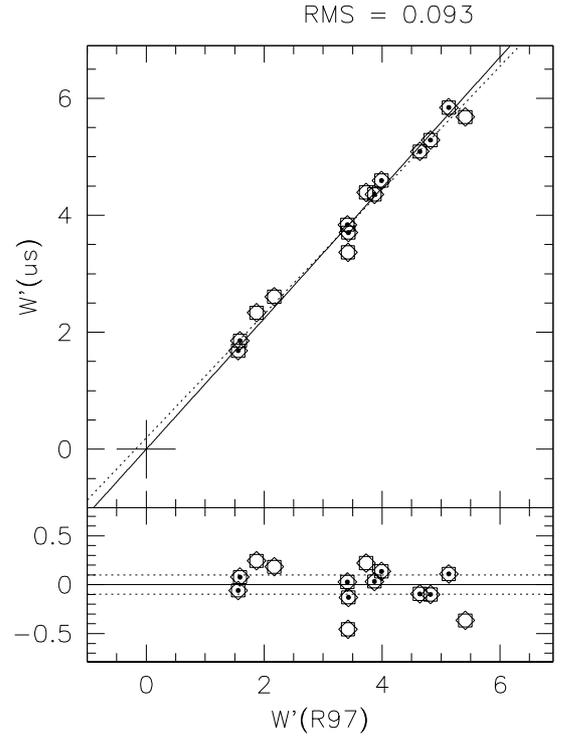} 
\par\end{centering}

\caption{In the upper panel reduced equivalent widths of calibration clusters
on our scale are plotted against those of R97. The data were fit with
a linear relation passing through $(0,0)$, and a $1\,\sigma$ rejection
was also applied after a preliminary fit. This leaves the clusters
represented by encircled dots. A free linear fit to the whole sample
is represented by the dotted line. In the lower panel the difference
of our $\W$ minus the converted R97 ones is shown, and the average
difference $\pm\sigma$ are shown by the solid and dotted lines, respectively.\label{fig:Reduced-equivalent-widths}}
\end{figure}

\begin{figure}
\begin{centering}
\includegraphics[width=0.9\columnwidth]{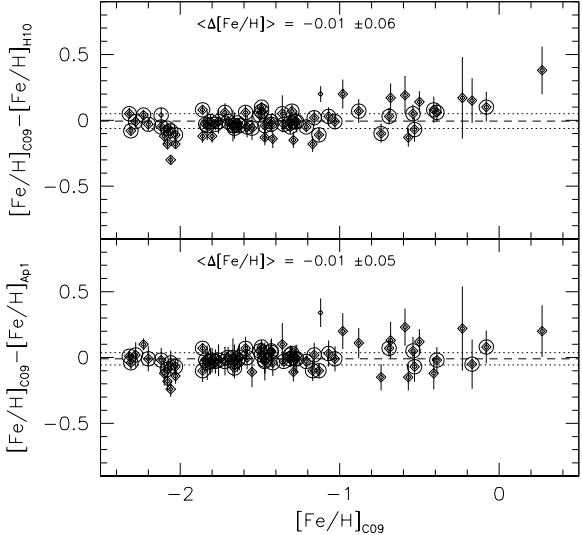} 
\par\end{centering}

\caption{Same as Fig. \ref{fig:Differences-in-metallicity} for R97 clusters.
Note that the ${\rm [Fe/H]}$ for the most metal-rich data point was
extrapolated beyond the range of our calibration. \label{fig:Differences-in-metallicity-r97}}
\end{figure}

To enlarge the sample of clusters with homogeneous metallicities,
we took the reduced equivalent widths from R97 and converted their
$\W$ values to our $\W$ scale. As Fig.~\ref{fig:Reduced-equivalent-widths}
shows, R97 widths can be converted to our scale by the simple relation
$\W_{{\rm G09}}=a\times\W_{{\rm R97}}$, where $a=1.117\pm0.043$,
which was obtained by comparing $\W$ of calibration clusters. For
these clusters, after a $1\sigma$ rejection, the average difference
$\pm$ sigma between our $\W$ and the converted R97 ones is $0.0003\pm0.0984$~dex
(lower panel of Fig.~\ref{fig:Reduced-equivalent-widths}). Apart
from calibration clusters, we have four objects in common with R97
(NGC2808, Rup106, NGC6715, and Terzan~7) which can be used to test
the transformation between the two systems. After excluding NGC6715
because of the metallicity dispersion, the comparison shows that for
the remaining three clusters our $\W$ are on average $0.15\pm0.18$~\AA{}\
larger than the converted R97 ones. This offset might seem large,
but on the other hand there are also two calibration clusters with
a similar deviation (see lower panel of Fig.~\ref{fig:Reduced-equivalent-widths}).
We therefore expect that a larger set of common clusters would yield
a null average difference between the G09 and the R97 converted $\W$
as for the calibration clusters.

Once the $\W$ of R97 were transformed onto our system, we could apply
the same calibration used to transform our $\W$ values to metallicity
values on the C09 scale. The new R97 metallicities together with {[}Fe/H{]}
from H10, and from Appendix 1 of C09, are given in Table~\ref{tab:All-[Fe/H]-values-r97}.
Figure~\ref{fig:Differences-in-metallicity-r97} shows a comparison
of our new {[}Fe/H{]} values to those of H10 and Appendix~1 of C09.
The same comparison for our program clusters was shown in Fig.~\ref{fig:Differences-in-metallicity},
and an inspection of the two figures reveals that the metallicity
differences in the case of R97 clusters show a scatter which is about
half that for our program clusters. This is not unexpected since the
R97 compilation of equivalent widths is at the base of all recent
abundances scales, so for those clusters the internal scatter is small.
In the future the twenty clusters studied here will enter the set
of objects with well-determined and homogeneous metallicities.

\subsection{Impact of the new abundances on the system of GGCs}

\begin{figure*}
\begin{centering}
\includegraphics[width=0.9\textwidth]{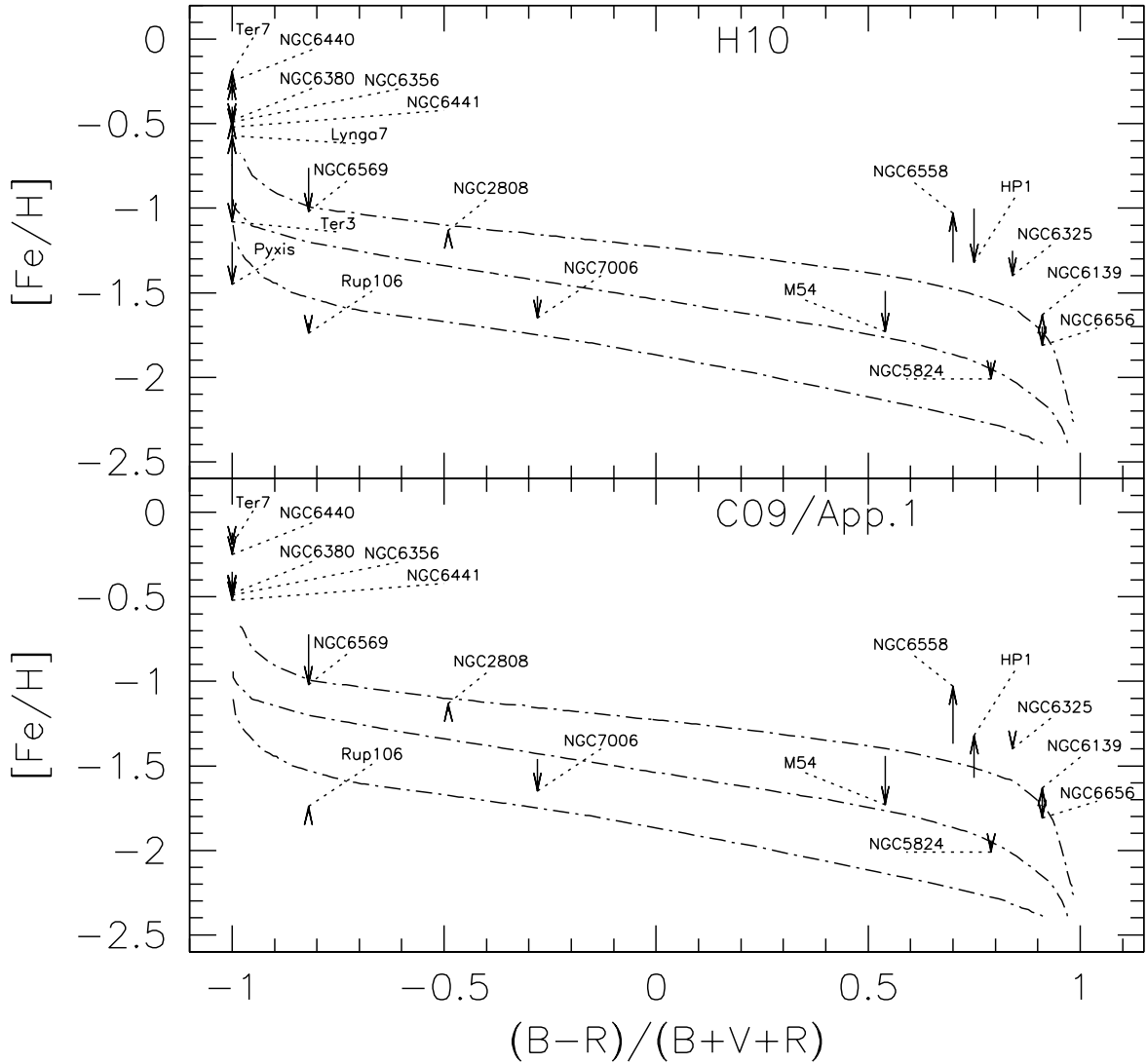} 
\par\end{centering}

\caption{The metallicity of our program clusters on the C09 scale is plotted
here against HB-type from Mackey and van den Bergh (\cite{mackey_vdb05}).
Isochrones are from Rey et al. (\cite{rey+01}), and are separated
by $1.1$ Gyr from top to bottom. The oldest isochrone gives the age
of clusters at $R<8$~kpc (Rey et al. 2001). Pal 7 is not present
in Mackey and van den Bergh (\cite{mackey_vdb05}), so it is not plotted
here. The arrows connect the position of the cluster if another {[}Fe/H{]}
estimate is used, to the position determined by our metallicity scale.
The other scale is H10 on the upper panel, and Appendix~1 of C09
in the lower panel. Consistent changes for both scales are those of
NGC7006, NGC6569, and NGC6715 which become younger, and that of NGC6558
which becomes older. Pyxis, Terzan~3, and Lynga~7 appear only in
the upper panel because they have no {[}Fe/H{]} estimates in C09.
Comparing our abundances to H10 only, Lynga~7 becomes older, while
the other two become younger. In fact Pyxis would now be as young
as Rup106. \label{fig:The-metallicity-of-hbtype}}
\end{figure*}

A common way to classify GGCs is to use the {[}Fe/H{]} vs. HB type
diagram, first introduced by Zinn (\cite{zinn93a}, Z93). Such diagram
for our clusters is shown in Fig.~\ref{fig:The-metallicity-of-hbtype},
where the HB type is taken from Mackey \& van den Bergh (\cite{mackey_vdb05}).
The figure shows that in 50\% of the cases our new metallicity values
do not change the clusters' classification significantly, but for
the other 50\% of objects the change in {[}Fe/H{]} implies a change
in age of $\sim1$~Gyr. In particular NGC7006, NGC6569 and NGC6715
move to younger ages, and NGC6558 moves to older ages, if our new
abundances are adopted. The cases of HP1 and NGC6380 are instead more
ambiguous, because their change in age depends on which comparison
abundances are used. With our new metallicity determination the location
of HP1 in the Z93 diagram moves in the direction of younger ages when
compared with its location using the H10 metallicity estimate, and
in the direction of older ages when compared to its location with
the C09/Appendix~1 abundance estimate. Our new metallicity for NGC6380
is similar to that of C09/Appendix 1 implying a null age change, but
the cluster becomes older if our new {[}Fe/H{]} is used instead of
that of H10. There are three clusters that do not have an entry in
C09/Appendix~1, so based on the change with respect to the H10 metallicity
only, our new {[}Fe/H{]} values would mean that both Pyxis and Terzan~3
become younger, while Lynga~7 becomes older.

In the classification of Z93 the division between so-called ``young
halo'' and ``old halo'' objects starts at an isochrone intermediate
between the oldest two of Fig.~\ref{fig:The-metallicity-of-hbtype}
(see, e.g, Fig.~5 of Mackey and Gilmore \cite{mackey_gilmore04}),
so the greatest impact of our new abundances is for NGC~6715 (M54)
which moves into the ``young halo'' area of the diagram. For clusters
with H10 metallicity only, Lynga~7 becomes ``old halo'' and Terzan~3
becomes ``young halo''. Pyxis was already young halo, but it becomes
the youngest cluster in our sample, together with Rup106.

Very recently Dotter et al. (\cite{dotter+11}) have provided age
estimations for six GGCs at Galactocentric distances $R_{{\rm GC}}>15$~kpc,
including three clusters of our sample. They found that Rup106, and
Pyxis are 1-2 Gyr younger than inner halo GCs with similar metallicities,
and that NGC7006 is marginally younger than its inner halo counterparts.
Figure~\ref{fig:The-metallicity-of-hbtype} is consistent with Dotter
et al. in terms of relative ages, but the ages predicted by Rey et
al. (\cite{rey+01}) isochrones are lower by $\sim1$~Gyr.

\section{Conclusions and outlook}

The main result of this work is presented in Table~\ref{tab:metallicities},
which gives the average metallicity on the C09 scale for our twenty
program clusters, most of which are globular clusters in the distant
halo or the Galactic bulge, with no or scanty spectroscopic observations.
Together with $68$ clusters in Table~\ref{tab:All-[Fe/H]-values-r97}
($72-4$ in common), they add up to $88$ clusters having metallicities
on the same scale, which is currently the largest set of objects with
{[}Fe/H{]} measured with the \emph{same method}. In addition Table~\ref{tab:Radial-Velocities}
gives also homogeneous radial velocities for our program clusters,
which can be important for kinematical studies of the GGC system.

A comparison with literature metallicities revealed that our clusters
had {[}Fe/H{]} values with uncertainties at the level of $0.2$~dex
RMS, because of inconsistent zero-points among different authors (Fig.~\ref{fig:Differences-in-metallicity}).
In particular the largest revisions of {[}Fe/H{]} values happen for
NGC6569, NGC6715, NGC6558, HP1, NGC6380, and NGC7006. When analyzed
in the light of the {[}Fe/H{]} vs. HB-type diagram, most clusters
appear younger with the new metallicities. For NGC~6715 (M54) the
younger age means that it now belongs to the ``young halo'' area
of the diagram.

For all clusters we measured metallicities for an average of $10$
stars, which let us search for metallicity dispersions. The M22 dispersion
was discussed in Da Costa et al. (2009), and we confirmed the {[}Fe/H{]}
dispersion of M54 with $15$ stars. In addition with $17$ stars measured,
we detected a probable dispersion also in NGC~5824.

Some of our targets were included as `special' objects, in particular
Terzan~7 as Sgr cluster, NGC7006 as second parameter object, and
NGC6325, NGC6356, NGC6440, NGC6441, and HP1 as bulge members. The
new Terzan~7 metallicity is not so different from that in H10 and
C09, so we confirm that it follows the age-metallicity relation of
the Sgr dwarf. NGC~7006 was the first cluster where a violation of
the HB-morphology vs. {[}Fe/H{]} relation was found, by Sandage \&
Wildey (\cite{sandage_wildey67}). Its HB is too red for its metallicity.
Our new metallicity is lower than literature values, so the violation
becomes even more severe, although in the framework of Z93 it would
be explained by an age lower than fiducial ``old halo'' clusters,
as shown by Fig.~\ref{fig:The-metallicity-of-hbtype}.

For bulge clusters our study, besides providing in many cases the
first spectroscopic metallicity measurements of individual stars,
allows a useful pre-selection of cluster members, to be followed up
with high-resolution studies. By comparing Tables~\ref{tab:Observations-log}
to \ref{tab:metallicities} we can see that only $20\%$ up to $58\%$
of stars observed in these clusters were confirmed as members, so
our work will indeed allow future campaigns to save substantial observing
time, not having to worry about foreground/background contamination.

Finally the detection of a possible metallicity dispersion in NGC5824
deserves confirmation with a larger database of stellar {[}Fe/H{]},
which we are in the process of collecting both at the VLT and Gemini.
Interestingly, clusters where dispersions have been detected are all
possible nuclear clusters of dwarf galaxies in different stages of
merging with the Milky Way, and they also have multiple stellar populations
(see, e.g., Piotto \cite{piotto09}). In this context, with our new
metallicity M54 now falls on the same low-age isochrone of NGC5824
in the Z93 diagram (see Fig.~\ref{fig:The-metallicity-of-hbtype}),
so this feature might become another distinctive trait of at least
some nuclear clusters. In this respect it is interesting to note the
work of Paudel et al. (\cite{paudel+11}): they determined the ages
of the stellar populations of 26 early-type dwarf galaxies in the
Virgo cluster, and they found that in most cases (21 galaxies) their
nuclei are younger than the main bodies.

We expect that our database of homogeneous metallicities for $\sim90$
clusters (and radial velocities for $20$ clusters) will be the foundation
of many future studies of age-metallicity and age-kinematics relations.
We plan to collect data on the 29 remaining clusters to complete our
project. 
\begin{acknowledgements}
We wish to thank L. Rizzi and Y. Momany for their help with the initial
observation design. We also thank the anonymous referee for suggestions
that improved the presentation of this paper. Part of this work was
completed with two visits of IS to the Padova Astronomical Observatory,
which hospitality is warmly acknowledged. The visits were supported
by grants of ESO's Director General. GDC also acknowledges the warm
hospitality of Padova Astronomical Observatory in supporting his two
visits, and the support from ESO for hosting a visit to ESO-Santiago
that enabled considerable progress in completing this paper. GDC's
research is supported in part by the Australian Research Council through
Discovery Projects grant DP0878137. EVH acknowledges financial support
through the PRIN INAF 2009 (P.I. R. Gratton). MG acknowledges financial
support from the Belgian Federal Science Policy (project MO/33/026).
BB acknowledges partial financial support by CNPq and FAPESP. IS wishes
to dedicate this paper to the memory of Carlo Izzo, whose creativity
mixed with logic and wit made an exquisite person and a dear friend.\end{acknowledgements}

\clearpage{}

\longtab{7}{ %
\begin{longtable}{llllllllll}
\caption{\label{tab:All-[Fe/H]-values-r97} Metallicities for globular clusters
from R97 on the Carretta et al. (2009) scale}
\endfirsthead
\hline 
\hline 
Cluster  & alt.  & $W'$  & $\sigma(W')$  & ${\rm [Fe/H]}_{{\rm C09}}$  & $\epsilon$  & ${\rm [Fe/H]}_{{\rm H10}}$  & H10wt  & ${\rm [Fe/H}_{{\rm C09}}^{{\rm Ap1}}$  & $\epsilon$\tabularnewline
\hline 
 &  &  &  &  &  &  &  &  & \tabularnewline
104  & 47 Tuc  & 5.06  & 0.07  & -0.69  & 0.06  & -0.72  & 10  & -0.76  & 0.02\tabularnewline
288  &  & 4.09  & 0.08  & -1.35  & 0.04  & -1.32  & 3  & -1.32  & 0.02\tabularnewline
362  &  & 4.16  & 0.09  & -1.31  & 0.05  & -1.26  & 5  & -1.30  & 0.04\tabularnewline
1261  &  & 4.21  & 0.11  & -1.28  & 0.06  & -1.27  & 3  & -1.27  & 0.08\tabularnewline
Eridanus  &  & 3.88  & 0.23  & -1.47  & 0.12  & -1.43  & 4  & -1.44  & 0.08\tabularnewline
1851  &  & 4.68  & 0.16  & -0.98  & 0.11  & -1.18  & 9  & -1.18  & 0.08\tabularnewline
1904  & M79  & 3.48  & 0.14  & -1.66  & 0.06  & -1.60  & 6  & -1.58  & 0.02\tabularnewline
2298  &  & 2.54  & 0.09  & -2.03  & 0.03  & -1.92  & 5  & -1.96  & 0.04\tabularnewline
2808  &  & 4.19  & 0.10  & -1.29  & 0.06  & -1.14  & 4  & -1.18  & 0.04\tabularnewline
Pal3  &  & 3.45  & 0.26  & -1.67  & 0.12  & -1.63  & 5  & -1.67  & 0.08\tabularnewline
3201  &  & 3.83  & 0.07  & -1.49  & 0.04  & -1.59  & 5  & -1.51  & 0.02\tabularnewline
Pal4  &  & 4.07  & 0.26  & -1.36  & 0.14  & -1.41  & 5  & -1.46  & 0.08\tabularnewline
4147  &  & 3.13  & 0.14  & -1.81  & 0.06  & -1.80  & 4  & -1.78  & 0.08\tabularnewline
4372  &  & 2.09  & 0.13  & -2.20  & 0.05  & -2.17  & 4  & -2.19  & 0.08\tabularnewline
Rup106  &  & 3.14  & 0.12  & -1.80  & 0.05  & -1.68  & 6  & -1.78  & 0.08\tabularnewline
4590  & M68  & 1.78  & 0.08  & -2.31  & 0.03  & -2.23  & 5  & -2.27  & 0.04\tabularnewline
4833  &  & 2.54  & 0.08  & -2.03  & 0.03  & -1.85  & 4  & -1.89  & 0.05\tabularnewline
5053  &  & 1.87  & 0.14  & -2.28  & 0.05  & -2.27  & 4  & -2.30  & 0.08\tabularnewline
5286  &  & 3.39  & 0.10  & -1.70  & 0.04  & -1.69  & 4  & -1.70  & 0.07\tabularnewline
5694  &  & 2.46  & 0.17  & -2.06  & 0.06  & -1.98  & 5  & -2.02  & 0.07\tabularnewline
5897  &  & 2.41  & 0.11  & -2.08  & 0.04  & -1.90  & 3  & -1.90  & 0.06\tabularnewline
5904  & M5  & 4.17  & 0.12  & -1.31  & 0.07  & -1.29  & 11  & -1.33  & 0.02\tabularnewline
5927  &  & 5.39  & 0.11  & -0.41  & 0.10  & -0.49  & 5  & -0.29  & 0.07\tabularnewline
5986  &  & 3.53  & 0.11  & -1.63  & 0.05  & -1.59  & 4  & -1.63  & 0.08\tabularnewline
Pal14  &  & 3.48  & 0.23  & -1.66  & 0.10  & -1.62  & 4  & -1.63  & 0.08\tabularnewline
6093  & M80  & 3.20  & 0.09  & -1.78  & 0.04  & -1.75  & 6  & -1.75  & 0.08\tabularnewline
6121  & M4  & 4.32  & 0.07  & -1.21  & 0.04  & -1.16  & 11  & -1.18  & 0.02\tabularnewline
6101  &  & 2.36  & 0.17  & -2.10  & 0.06  & -1.98  & 3  & -1.98  & 0.07\tabularnewline
6144  &  & 2.47  & 0.08  & -2.06  & 0.03  & -1.76  & 4  & -1.82  & 0.05\tabularnewline
6171  & M107  & 4.46  & 0.08  & -1.13  & 0.05  & -1.02  & 4  & -1.03  & 0.02\tabularnewline
6205  & M13  & 3.65  & 0.17  & -1.58  & 0.08  & -1.53  & 13  & -1.58  & 0.04\tabularnewline
6218  & M12  & 4.18  & 0.09  & -1.30  & 0.05  & -1.37  & 5  & -1.33  & 0.02\tabularnewline
6235  &  & 3.96  & 0.13  & -1.42  & 0.07  & -1.28  & 4  & -1.38  & 0.07\tabularnewline
6254  & M10  & 3.82  & 0.09  & -1.49  & 0.05  & -1.56  & 4  & -1.57  & 0.02\tabularnewline
Pal15  &  & 2.29  & 0.15  & -2.12  & 0.05  & -2.07  & 4  & -2.10  & 0.08\tabularnewline
6266  & M62  & 4.41  & 0.09  & -1.16  & 0.06  & -1.18  & 3  & -1.18  & 0.07\tabularnewline
6273  & M19  & 3.01  & 0.12  & -1.86  & 0.05  & -1.74  & 4  & -1.76  & 0.07\tabularnewline
6304  &  & 5.41  & 0.08  & -0.39  & 0.07  & -0.45  & 4  & -0.37  & 0.07\tabularnewline
6352  &  & 5.29  & 0.09  & -0.50  & 0.08  & -0.64  & 5  & -0.62  & 0.05\tabularnewline
6366  &  & 5.24  & 0.08  & -0.54  & 0.07  & -0.59  & 3  & -0.59  & 0.08\tabularnewline
6362  &  & 4.39  & 0.09  & -1.17  & 0.06  & -0.99  & 4  & -1.07  & 0.05\tabularnewline
6397  &  & 2.42  & 0.08  & -2.08  & 0.03  & -2.02  & 7  & -1.99  & 0.02\tabularnewline
6496  &  & 5.25  & 0.10  & -0.53  & 0.09  & -0.46  & 3  & -0.46  & 0.07\tabularnewline
6522  &  & 3.88  & 0.11  & -1.47  & 0.06  & -1.34  & 4  & -1.45  & 0.08\tabularnewline
6535  &  & 3.06  & 0.28  & -1.83  & 0.11  & -1.79  & 3  & -1.79  & 0.07\tabularnewline
6528  &  & 6.05  & 0.15  & 0.27  & 0.18  & -0.11  & 7  & 0.07  & 0.08\tabularnewline
6544  &  & 3.94  & 0.11  & -1.43  & 0.06  & -1.40  & 4  & -1.47  & 0.07\tabularnewline
6541  &  & 3.04  & 0.08  & -1.84  & 0.03  & -1.81  & 4  & -1.82  & 0.08\tabularnewline
6553  &  & 5.73  & 0.11  & -0.08  & 0.11  & -0.18  & 6  & -0.16  & 0.06\tabularnewline
6624  &  & 5.21  & 0.08  & -0.57  & 0.07  & -0.44  & 4  & -0.42  & 0.07\tabularnewline
6626  &  & 4.48  & 0.10  & -1.12  & 0.06  & -1.32  & 2  & -1.46  & 0.09\tabularnewline
6638  &  & 4.82  & 0.12  & -0.88  & 0.09  & -0.95  & 4  & -0.99  & 0.07\tabularnewline
6637  & M69  & 5.01  & 0.09  & -0.74  & 0.07  & -0.64  & 4  & -0.59  & 0.07\tabularnewline
6681  & M70  & 3.51  & 0.08  & -1.64  & 0.04  & -1.62  & 3  & -1.62  & 0.08\tabularnewline
6712  &  & 4.60  & 0.09  & -1.03  & 0.06  & -1.02  & 4  & -1.02  & 0.07\tabularnewline
6715  & M54  & 3.71  & 0.19  & -1.55  & 0.09  & -1.49  & 7  & -1.44  & 0.07\tabularnewline
6717  & Pal9  & 4.22  & 0.13  & -1.27  & 0.08  & -1.26  & 3  & -1.26  & 0.07\tabularnewline
6723  &  & 4.55  & 0.10  & -1.07  & 0.07  & -1.10  & 3  & -1.10  & 0.07\tabularnewline
6752  &  & 3.81  & 0.07  & -1.50  & 0.04  & -1.54  & 8  & -1.55  & 0.01\tabularnewline
Ter7  &  & 5.64  & 0.17  & -0.17  & 0.17  & -0.32  & 6  & -0.12  & 0.08\tabularnewline
Arp2  &  & 3.24  & 0.22  & -1.76  & 0.09  & -1.75  & 5  & -1.74  & 0.08\tabularnewline
6809  & M55  & 3.01  & 0.08  & -1.86  & 0.03  & -1.94  & 4  & -1.93  & 0.02\tabularnewline
Ter8  &  & 2.31  & 0.14  & -2.12  & 0.05  & -2.16  & 2  & \ldots{}  & \ldots{}\tabularnewline
Pal11  &  & 5.58  & 0.31  & -0.23  & 0.31  & -0.40  & 4  & -0.45  & 0.08\tabularnewline
6838  & M71  & 5.18  & 0.17  & -0.59  & 0.14  & -0.78  & 11  & -0.82  & 0.02\tabularnewline
6981  & M72  & 3.94  & 0.12  & -1.43  & 0.06  & -1.42  & 4  & -1.48  & 0.07\tabularnewline
7078  & M15  & 1.74  & 0.12  & -2.32  & 0.04  & -2.37  & 11  & -2.33  & 0.02\tabularnewline
7089  & M2  & 3.63  & 0.10  & -1.59  & 0.05  & -1.65  & 5  & -1.66  & 0.07\tabularnewline
7099  & M30  & 2.00  & 0.10  & -2.23  & 0.04  & -2.27  & 4  & -2.33  & 0.02\tabularnewline
Pal12  &  & 5.08  & 0.13  & -0.68  & 0.11  & -0.85  & 6  & -0.81  & 0.08\tabularnewline
7492  &  & 3.33  & 0.17  & -1.72  & 0.07  & -1.78  & 5  & -1.69  & 0.08\tabularnewline
 &  &  &  &  &  &  &  &  & \tabularnewline
\hline 
\end{longtable}}

\appendix

\section{Transformation between the AD91 and G09 measurements\label{sec:Transformation-between-the}}

\begin{figure}
\begin{centering}
\includegraphics[angle=-90,width=0.9\columnwidth]{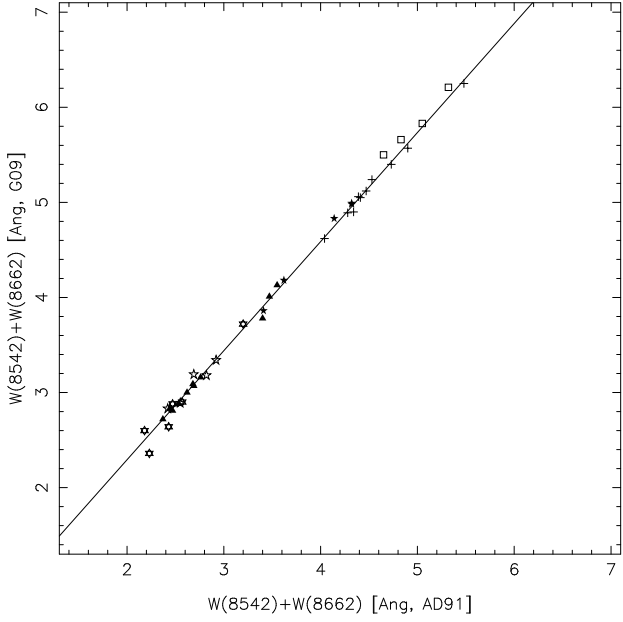} 
\par\end{centering}

\caption{The sum of equivalent widths for the strongest two CaT lines measured
with Gaussian-only fits are plotted here against the same quantity
measured with Gaussian+Lorentzian fits. The latter are taken from
the G09 paper, for stars in the clusters M5 (+ signs), NGC4372 (filled
triangles), NGC6752 (filled 5-point stars), NGC6397 (open 5-point
stars), NGC 6171 (open squares) and NGC 4590 (6-point stars). The
interpolating line has slope $f=1.147$, and it was obtained by minimizing
residuals and imposing the passage through the origin. \label{fig:The-sum-of-2003}}
\end{figure}

\begin{figure}
\begin{centering}
\includegraphics[angle=-90,width=0.9\columnwidth]{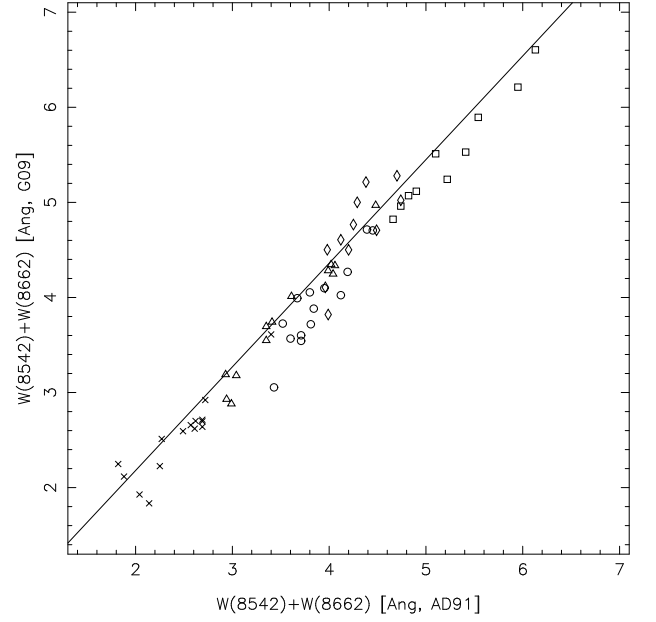} 
\par\end{centering}

\caption{Same as Fig.~\ref{fig:The-sum-of-2003}, for the stars observed for
this work. The symbols identify NGC3201 (open circles), M10 (open
triangles), M4 (open diamonds), M15 (x signs) and M71 (open squares).
The fitting line has slope $f=1.090$ and passes through the origin.
\label{fig:The-sum-of-2006}}
\end{figure}

We define the reduced equivalent width for a single star as $W^{\prime}=S-a\,(V-V_{{\rm HB}})$
where $S=W_{8542}+W_{8662}$%
\footnote{We use $S$ here instead of $\Sigma W$ because in this way the formulas
are easier to read.%
}. So we can write for the (G)09 and (A)D91 systems:

\[
W_{{\rm G}}^{\prime}=S_{{\rm G}}-a_{{\rm G}}\,(V-V_{{\rm HB}})
\]

\[
W_{{\rm A}}^{\prime}=S_{{\rm A}}-a_{{\rm A}}\,(V-V_{{\rm HB}})
\]
 And the averaged reduced equivalent widths are:

\[
\left\langle W_{{\rm G}}^{\prime}\right\rangle =[\Sigma S_{{\rm G}}-a_{{\rm G}}\,\Sigma(V-V_{{\rm HB}})]/N
\]

\[
\left\langle W_{{\rm A}}^{\prime}\right\rangle =[\Sigma S_{{\rm A}}-a_{{\rm A}}\,\Sigma(V-V_{{\rm HB}})]/N
\]
 By comparing measurements of individual $S$ made with the G09 or
the AD91 method, we found that $S_{{\rm G}}=f\, S_{{\rm A}}$, and
by imposing a linear fit that is passing through $(0,0)$ we found
for clusters of the Leo~I paper $f=1.147$ (see Fig.~\ref{fig:The-sum-of-2003})
and for clusters in this run $f=1.090$ (see Fig.~\ref{fig:The-sum-of-2006}).
So then in the G09 system:

\[
\left\langle W_{{\rm G}}^{\prime}\right\rangle =[f\,\Sigma S_{{\rm A}}-a_{{\rm G}}\,\Sigma(V-V_{{\rm HB}})]/N
\]
 Then we have

\[
\Sigma S_{{\rm A}}=N\left\langle W_{{\rm A}}^{\prime}\right\rangle +a_{{\rm A}}\,\Sigma(V-V_{{\rm HB}})
\]
 And with a substitution

\[
\left\langle W_{{\rm G}}^{\prime}\right\rangle =[f\,(N\left\langle W_{{\rm A}}^{\prime}\right\rangle +a_{{\rm A}}\,\Sigma_{{\rm HB}})-a_{{\rm G}}\,\Sigma_{{\rm HB}}]/N
\]
 where for simplicity we wrote $\Sigma_{{\rm HB}}=\Sigma(V-V_{{\rm HB}})$.
Rearranging and simplifying a bit more:

\[
N\left\langle W_{{\rm G}}^{\prime}\right\rangle =f\, N\left\langle W_{{\rm A}}^{\prime}\right\rangle +(f\, a_{{\rm A}}-a_{{\rm G}})\Sigma_{{\rm HB}}
\]
 Which gives $\left\langle W_{{\rm G}}^{\prime}\right\rangle $ as
a function of $f$ (with $f=1.090$ or $f=1.147$), the number of
stars $N$ measured for each cluster, the averaged reduced equivalent
width in the AD91 system, the slopes $a_{{\rm A}}$ and $a_{{\rm G}}$
of the $S$ vs. $V-V_{{\rm HB}}$ relations in the two systems, and
the sum of the magnitude differences with respect to the HB, $\Sigma_{{\rm HB}}$.
The difference in slopes $f$ is probably due to the fact that older
spectra had a better resolution (slit of $0\farcs8$ instead of $1\arcsec$),
so it might be that the Gauss-only fit does a worse job with line
wings in that case. 
\end{document}